\documentclass[pdflatex,sn-mathphys-num]{sn-jnl}

\DeclareTextCommandDefault{\k}[1]{{\fontencoding{T1}\selectfont\k{#1}}}
\usepackage{graphicx}
\usepackage{amsmath,amssymb,amsfonts}
\usepackage{amsthm}
\usepackage{xcolor}
\usepackage{textcomp}
\usepackage{manyfoot}
\usepackage{booktabs}
\usepackage{algorithm}
\usepackage{algpseudocode}
\usepackage{placeins}

\theoremstyle{thmstyleone}
\newtheorem{theorem}{Theorem}
\newtheorem{lemma}[theorem]{Lemma}
\newtheorem{proposition}[theorem]{Proposition}
\newtheorem{corollary}[theorem]{Corollary}
\theoremstyle{thmstyletwo}
\newtheorem{example}{Example}
\newtheorem{remark}{Remark}
\theoremstyle{thmstylethree}
\newtheorem{definition}{Definition}

\newcommand{\N}{\mathbb{N}}
\newcommand{\Z}{\mathbb{Z}}
\newcommand{\cE}{\mathcal{E}}
\newcommand{\cB}{\mathcal{B}}
\newcommand{\e}{\mathbf{e}}
\newcommand{\out}{\mathrm{out}}
\newcommand{\inn}{\mathrm{in}}
\newcommand{\IN}{\mathrm{IN}}
\newcommand{\OUT}{\mathrm{OUT}}
\newcommand{\FIX}{\textup{\textsc{Fix}}}
\newcommand{\FREE}{\textup{\textsc{Free}}}
\newcommand{\SFIX}{\textup{\textsc{Strict-Fix}}}
\newcommand{\SFREE}{\textup{\textsc{Strict-Free}}}
\newcommand{\abs}[1]{\lvert #1\rvert}
\newcommand{\floor}[1]{\lfloor #1\rfloor}
\newcommand{\YES}{\texttt{YES}}
\newcommand{\NO}{\texttt{NO}}

\graphicspath{{figures/}}
\makeatletter
\def\input@path{{sections/}}
\makeatother

\begin{document}

\title[Exact edge cover by temporal walks]{Exact edge cover by strict and non-strict temporal walks}

\author[1]{\fnm{Duc Thien} \sur{Quach}}\email{23020161@vnu.edu.vn}

\author*[1]{\fnm{Manh Cuong} \sur{Tran}}\email{tranmanhcuong@vnu.edu.vn}

\affil[1]{\orgdiv{Faculty of Information Technology}, \orgname{VNU University of Engineering and Technology}, \orgaddress{\city{Hanoi}, \country{Vietnam}}}

\abstract{Exact edge cover asks whether the edges of a temporal graph with $n$ vertices and $m$ edges can be partitioned into time-respecting walks with a prescribed number of walks starting at each vertex and, in the fixed-end variant, ending at each vertex. For non-strict walks, we prove that the problem is polynomially solvable by a new approach based on M-convexity, which also closes a gap in the proof sketch of an earlier static reduction. An orientation of the edges is realizable by walks exactly when walk counters stay non-negative and every snapshot component contains a walk, and only components with all degrees and counters even can violate the latter condition. Once a dead state is removed, the states of each such component form an M-convex set. This yields an augmenting path algorithm that returns walks or an infeasibility certificate, using at most $m/2$ augmentations and $O(n+nm^2)$ arithmetic operations with path-reversing re-orientation. The implementation was validated on small instances. For strict walks, we prove that the problem is NP-complete, already with six time labels for prescribed ends and two for free ends.}

\keywords{Temporal graph, Exact edge cover, Temporal walk, M-convex set, Submodular flow, NP-completeness}

\pacs[MSC Classification]{05C85, 90C27, 90C35, 68Q17}

\maketitle

\section{Introduction}\label{sec:intro}

Temporal graphs model connections whose availability changes with time~\cite{holme2012,michail2016}. In arc routing, edges are tasks, while vehicle scheduling assembles time-ordered trips into duties~\cite{edmonds1973,corberan2015,bunte2009}. These settings motivate covering edges by walks that respect their time labels. We consider undirected edges, choosing their traversal direction together with the walks. Experiments ignore recorded directions and travel times, testing correctness and scalability rather than a routing application.

An exact edge cover partitions temporal edges among walks, traversing each edge once. The input prescribes start counts and, in \FIX{}, end counts without pairing individual starts and ends. In \FREE{}, the solution chooses the end vertices. Empty walks are allowed. A non-strict walk uses non-decreasing time labels and may traverse several edges of one snapshot, whereas a strict walk uses strictly increasing labels and at most one edge per snapshot. Equal time stamps occur in coarsely recorded contact data, and the strict/non-strict distinction can change complexity~\cite{casteigts2024,erlebach2023}, including for temporal exploration.

Deligkas et al.~\cite{deligkas2025} introduced temporal exact edge cover at AAAI 2025, establishing the problem and its complexity results. This paper builds on their problem and notation. They claimed polynomial solvability for non-strict walks through a static reduction. Their NP-completeness result without prescribed starts concerns minimizing walk count. Strict walks with prescribed terminals were left open. The proof sketch of their Theorem~6.3 misses a temporal availability condition: every snapshot component containing an edge must contain a walk immediately before its time, which static connectivity to a terminal does not ensure. Section~\ref{sec:counterexample} gives a five-vertex instance whose static graph has an exact edge cover but whose temporal graph does not, under the end-terminal construction in Figure~3 of~\cite{deligkas2025}. This gap does not refute their claim. We prove the non-strict polynomial result differently and settle the strict variants.

Our method characterizes realizable orientations by non-negative walk counts and a walk in every snapshot component. Only components with even degrees and counters can violate the latter. Pair counting exposes M-convex component states, and removing the dead state preserves M-convexity. This yields an M-convex intersection formulation and an augmenting path algorithm that repairs dead components or certifies infeasibility. To our knowledge, dead-state treatment is new. 

\paragraph{Contributions.}
\begin{enumerate}
\item \emph{Characterization and gap.} An orientation is realizable if and only if the counting condition holds and every snapshot component contains a walk (Theorem~\ref{thm:char}). Only even components can violate the latter (Lemma~\ref{lem:noneven}), and instances without them reduce to one maximum flow (Corollary~\ref{cor:flow}). Under the end-terminal construction shown in Figure~3 of~\cite{deligkas2025}, a five-vertex instance shows a gap in the static reduction (Proposition~\ref{prop:aaai}, Lemma~\ref{lem:aaai}).
\item \emph{M-convex feasibility algorithm.} We prove that removing the dead state preserves M-convexity for even-component states (Lemma~\ref{lem:gamma}(b)), yielding a polynomial algorithm via feasibility of an M-convex intersection (Theorem~\ref{thm:reduction}). It returns walks or an infeasibility certificate (Theorem~\ref{thm:correct}) and has the bound of Theorem~\ref{thm:time}.
\item \emph{Strict-walk hardness.} The strict problem is NP-complete with prescribed ends and six time labels, and with free ends and two time labels, empty walks allowed (Theorem~\ref{thm:strict}).
\end{enumerate}

\section{Problem formulation and related work}\label{sec:prelim}

\subsection{Problem formulation}\label{sec:problem}

We write $\N=\{0,1,2,\ldots\}$ and use $\Z$ for the integers. An undirected temporal graph is $G=(V,\cE,\tau)$, with edge labels $\tau:\cE\to\N$, $n=\abs{V}$ and $m=\abs{\cE}$. Parallel edges are distinct. Loops are excluded. At time $t$, $G_t=(V,E_t)$, where $E_t=\{e\in\cE:\tau(e)=t\}$, is the \emph{snapshot}. A snapshot component contains at least one edge.

A \emph{non-strict temporal walk} is $W=(v_0,e_1,v_1,\ldots,e_\ell,v_\ell)$, where $e_i$ joins $v_{i-1}$ to $v_i$ and $\tau(e_1)\le\cdots\le\tau(e_\ell)$. Its start and end are $v_0,v_\ell$. An empty walk has $\ell=0$. Exact edge cover traverses each edge once across all walks.

The decision problems prescribe a vector $a\in\N^V$ of start counts, with $k=\sum_v a_v$.

\begin{definition}\label{def:problems}
\begin{itemize}
\item \FREE{}: decide whether there exist $k$ non-strict temporal walks that collectively traverse every edge of $\cE$ exactly once, with exactly $a_v$ walks starting at each vertex $v$.
\item \FIX{}: given an additional vector $z\in\N^V$, decide whether there exist walks satisfying \FREE{} with exactly $z_v$ walks ending at each vertex $v$.
\end{itemize}
\end{definition}

In \FIX{}, $a$ and $z$ prescribe endpoint counts without pairing starts to ends. Walks are non-strict unless stated otherwise.

Running times count unit-cost arithmetic operations on machine-word integers, including $a_v,z_v$. We return a \emph{compressed} solution (non-empty walks and each vertex's empty-walk count), since binary multiplicities can require exponentially many listed empty walks.

\subsection{Related work}\label{sec:related}

\paragraph{Exact edge cover in temporal graphs.}
Deligkas et al.~\cite{deligkas2025} study temporal paths, trails and walks (theorem and page numbers refer to the AAAI version, while the extended version is cited as arXiv v1). For strict walks with fixed terminals, Corollary~6.4 handles proper temporal graphs (each snapshot is a matching), and Theorem~6.5 gives an $O(2^k k\,t_{\max})$ algorithm when $|E_t|\in\{0,k\}$ ($t_{\max}$ is the largest time label). Directed cases are polynomial (Theorems~6.1 and~5.2). Their Theorem~6.3 mentions start terminals, while its proof and Figure~3 use end terminals. The extended version~\cite{deligkas2024arxiv} states both. Section~\ref{sec:counterexample} analyses in detail only the end-terminal construction shown in the proof and Figure~3. The variant with start terminals only is not addressed here.

\paragraph{Temporal networks and covering problems.}
Time-copy representations appear in~\cite{akrida2019}. Marino and Silva study a single walk that traverses every edge once in a temporal graph~\cite{marino2023}. For $k=1$ with a prescribed start and one label per edge, non-strict feasibility reduces, by Theorem~\ref{thm:char}, to a snapshot-by-snapshot Euler trail check, so we do not compare complexity results across the two models. Other related problems explore all edges~\cite{bumpus2023} or cover vertices of a temporal DAG~\cite{chakraborty2024}.

\paragraph{Discrete convexity and orientations.}
We use M-convex sets~\cite{murota1996,murota2003}, submodular flows~\cite{edmonds1977,franktardos1988,murota2003,fujishige2005}, degree-constrained orientations~\cite{hakimi1965,frank1980,frank2011} and augmenting paths~\cite{cunningham1985,fujishige2005}. Our additional constraint removes an even component's dead state. Related structures include strongly connected orientations~\cite{abdi2025}.

\section{Orientations, even components and a gap in the static reduction}\label{sec:char}

\subsection{Orientations and walk counters}

A walk that traverses the edge $\{u,v\}$ from $u$ to $v$ \emph{orients} it as $u\to v$, and $u$ is called the \emph{tail}. A solution thus induces an orientation of all edges. For an orientation, let $\out_t(v)$ and $\inn_t(v)$ be the numbers of arcs of $E_t$ with tail $v$ and with head $v$. The \emph{walk counter} $c_t(v)$ is the number of walks at $v$ just before time $t$. It starts at $a_v$ and evolves as
\[
  c_{t^+}(v)=c_t(v)-\out_t(v)+\inn_t(v),
\]
where $t^+$ denotes the moment just after $t$. Its parity is fixed by $a_v$ plus the number of edges incident to $v$ before $t$, independently of the orientation. A walk is counted once at its current vertex. At one time label it may traverse several successive edges.

Let $t_{v,1}<\dots<t_{v,q_v}$ be the time labels at which $v$ has an incident edge (the \emph{occurrences} of $v$), let $g_{v,j}$ be the degree of $v$ in $E_{t_{v,j}}$, and let $G_{v,j}=\sum_{i\le j}g_{v,i}$ and $G_v=G_{v,q_v}$. For an orientation, let $b_{v,j}$ be the number of arcs of $E_{t_{v,j}}$ with tail $v$ and $B_{v,j}=\sum_{i\le j}b_{v,i}$, where empty sums are zero. In \FIX{} we set $h_v=(a_v+G_v-z_v)/2$, the number of arcs with tail $v$ in every solution, as determined by the final count $z_v$.

\begin{theorem}\label{thm:char}
An orientation is realizable by walks that start according to $a$ (problem \FREE{}) if and only if, at every time $t$,
\begin{enumerate}
\item[(i)] $\out_t(v)-\inn_t(v)\le c_t(v)$ for every vertex $v$, and
\item[(ii)] every snapshot component $K$ of $E_t$ contains a walk, that is, $\sum_{v\in K}c_t(v)\ge 1$.
\end{enumerate}
For \FIX{} the additional condition is that the final number of walks at $v$ equals $z_v$ for every $v$, that is, $B_{v,q_v}=h_v$.
\end{theorem}

\begin{figure}[!htb]
\centering
\includegraphics{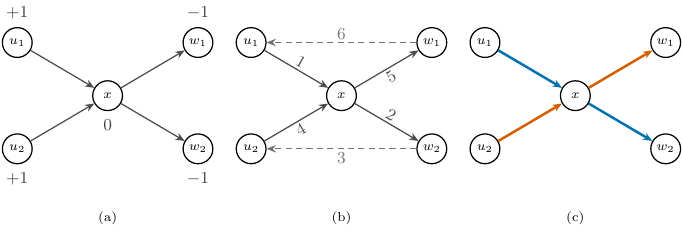}
\caption{Dummy arcs in the proof of Theorem~\ref{thm:char}, illustrated on a component with two surplus vertices $u_1,u_2$ and two deficit vertices $w_1,w_2$. (a) The oriented component, with $\out_t-\inn_t$ shown at each vertex. (b) Dashed dummy arcs join each deficit vertex to a surplus vertex. The numbers 1--6 indicate an Euler circuit. (c) Deleting the dummy arcs leaves two trails, from $u_1$ to $w_2$ and from $u_2$ to $w_1$.}
\label{fig:dummy}
\end{figure}

\begin{proof}
\emph{Necessity.} At time $t$, each edge that leaves $v$ is either the first edge that a walk uses at time $t$, and then the walk was at $v$ before $t$, so there are at most $c_t(v)$ such edges, or it follows an edge that enters $v$ at time $t$ on the same walk, and each entering edge has at most one successor. Hence $\out_t(v)\le c_t(v)+\inn_t(v)$, which is~(i). The edges that one walk uses at time $t$ form a trail inside one component $K$ of $E_t$, and the walk is located at the first vertex of this trail just before $t$. Hence every component that has edges contains a walk. In \FIX{} the final number of walks at $v$ is $a_v+G_v-2B_{v,q_v}=z_v$.

\emph{Sufficiency.} We process the times in increasing order and the components of $E_t$ independently. Let $\vec K$ be the directed graph of $K$ and let $r=\sum_{v\in K}\max(0,\out_t(v)-\inn_t(v))$. If $r>0$, add $r$ dummy arcs from vertices with $\inn>\out$ to vertices with $\out>\inn$, pairing each unit of deficit with one unit of surplus. The totals agree because the sum of $\out_t-\inn_t$ on $K$ is zero. This makes each vertex balanced. The augmented graph is weakly connected because $K$ is connected and the added arcs join vertices of $K$. Hence it has an Euler circuit. In cyclic order, deleting the $r$ dummy arcs leaves exactly $r$ trails (Fig.~\ref{fig:dummy}), each starting at the head of a dummy arc (a surplus vertex) and ending at the tail of the next dummy arc. A vertex $v$ with surplus is the head of exactly $\out_t(v)-\inn_t(v)$ dummy arcs, so that many trails start there. By~(i), $v$ holds enough walks to take these trails. If $r=0$, then $\vec K$ is Eulerian. By~(ii) some vertex of $K$ holds a walk, and this walk follows an Euler circuit that starts at that vertex. All other walks wait. The number of walks at every vertex after $t$ equals $c_{t^+}$, so the induction continues. In \FIX{} the final numbers are $z$.
\end{proof}

The proof of sufficiency yields an $O(n+m)$ algorithm (after sorting the edges) that constructs the walks from an orientation satisfying (i) and (ii) using Hierholzer's algorithm~\cite{hierholzer1873}. Given edge directions, \FREE{} and \FIX{} are thus decided in $O(n+m\log(m+1))$ time.

\subsection{Even components and the chain network}

\begin{definition}\label{def:even}
A snapshot component $K$ of $E_t$ is \emph{even} if every vertex of $K$ has even degree in $E_t$ and even walk counter $c_t$. An even component is \emph{dead} under an orientation if $c_t(v)=0$ for all $v\in K$.
\end{definition}

A pair of parallel edges and a triangle with even walk counters are even components. A path is not.

\begin{lemma}\label{lem:noneven}
If an orientation satisfies~(i) and the snapshot component $K$ of $E_t$ is not even, then $K$ satisfies~(ii).
\end{lemma}

\begin{proof}
If some vertex of $K$ has odd degree in $E_t$, then $\out_t-\inn_t$ is odd and therefore non-zero at that vertex. Since $\out_t-\inn_t$ sums to $0$ over $K$, some vertex has $\out_t-\inn_t>0$, and by~(i) it has $c_t\ge1$. If some counter $c_t(v)$ is odd, then it is a nonnegative odd integer, and hence is at least one: $c_t(v)$ equals $a_v\ge0$ at the first occurrence of $v$ and otherwise the counter $c_{t'^+}(v)$ after the previous occurrence $t'$, which is nonnegative by~(i) at $t'$.
\end{proof}

Condition~(i) requires the walk counter after the $j$-th occurrence of $v$, namely $a_v+G_{v,j}-2B_{v,j}$, to be non-negative, which is equivalent to $B_{v,j}\le\floor{(a_v+G_{v,j})/2}$.

We impose these bounds as capacities in a \emph{chain network} with a source and a sink $r$. For every occurrence $(v,j)$ there is a node $s_{v,j}$, and these nodes form a chain $s_{v,1}\to\dots\to s_{v,q_v}\to r$. The arc that leaves $s_{v,j}$ has capacity $C_{v,j}=\floor{(a_v+G_{v,j})/2}$ in \FREE{} and $C_{v,j}=\min(\floor{(a_v+G_{v,j})/2},h_v)$ in \FIX{}. For every edge $e=\{u,v\}$ with time label $t$ there is a node that receives one unit from the source and has two arcs of capacity $1$ to the occurrence nodes of $u$ and $v$ at time $t$.

The unit of $e$ goes to the endpoint chosen as its tail, so the flow on the arc that leaves $s_{v,j}$ equals $B_{v,j}$. The network has $O(m)$ nodes and arcs and, for \FREE{}, a flow of value $m$ corresponds exactly to an orientation that satisfies~(i). The counter after occurrence $j$ is $a_v+G_{v,j}-2B_{v,j}$, so the capacity on each outgoing chain arc enforces its nonnegativity.

\begin{figure}[t]
\centering
\includegraphics{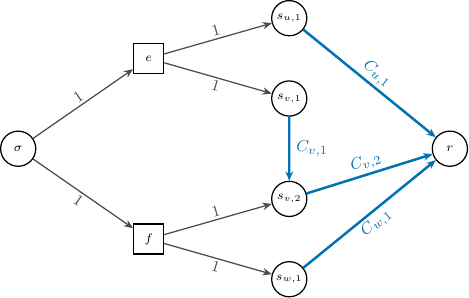}
\caption{Chain network for two successive temporal edges.}
\label{fig:chain}
\end{figure}

Figure~\ref{fig:chain} shows $e=\{u,v\}$ followed by $f=\{v,w\}$: units from $\sigma$ choose edge tails, and the thick chain arcs have capacities $C_{v,j}$.

\begin{corollary}\label{cor:flow}
If no snapshot has an even component, then \FREE{} and \FIX{} are decided by one maximum flow in the chain network. On every instance, a maximum flow of value less than $m$ certifies infeasibility.
\end{corollary}

\begin{proof}
By Theorem~\ref{thm:char} and Lemma~\ref{lem:noneven}, an orientation is realizable when there is no even component if and only if it satisfies~(i). For \FIX{}, after checking $\sum_va_v=\sum_vz_v$ and $h_v\in\N$ for all $v$, we have $\sum_vh_v=m$, so a flow of value $m$ saturates every last chain arc, which gives $B_{v,q_v}=h_v$. The second claim holds because~(i) is necessary.
\end{proof}

Proper temporal graphs have no even components, and strict and non-strict walks coincide on them, so Corollary~\ref{cor:flow} recovers Corollary~6.4 of~\cite{deligkas2025} with one maximum flow and extends it to \FREE{}.

\subsection{A gap in the static reduction}\label{sec:counterexample}

\begin{figure}[!htbp]
\centering
\includegraphics{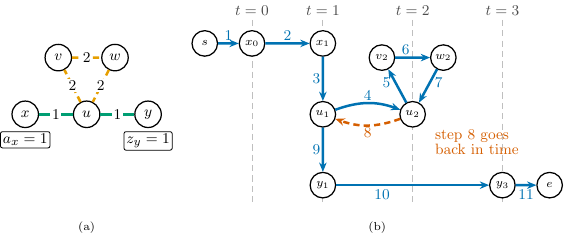}
\caption{A temporal instance (a) and a covering walk of its static reduction (b).}
\label{fig:aaai}
\end{figure}

\paragraph{The reduction.} For a vertex $v$, let $\delta_{\le t}(v)$ and $\delta_{>t}(v)$ count its incident edges with time labels at most $t$ and greater than $t$, respectively. A superscript $G$ restricts these counts to edges of the temporal graph $G$, without terminal edges. Deligkas et al.\ expand every vertex $v$ into copies $v_t$, one for each time at which $v$ has an edge, with a start terminal $s$ and an end terminal $e$. Under the end-terminal construction shown in Figure~3 of~\cite{deligkas2025}, the start-terminal edges attach to an additional copy at time $0$ and the end-terminal edges to an additional copy after the last time label (the copies $x_0$ and $y_3$ in Fig.~\ref{fig:aaai}(b)). Consecutive copies are joined by $\alpha=\min(\delta_{\le t}(v),\delta_{>t}(v))$ edges, where, for the end-terminal construction shown in the proof of Theorem~6.3 of~\cite{deligkas2025}, $\delta$ counts the edges of the static graph $S$ including the $a_v$ start-terminal and $z_v$ end-terminal edges at $v$ (so $\alpha=\min(a_v+\delta^G_{\le t}(v),\,z_v+\delta^G_{>t}(v))$ in the notation below), of which $\lceil\alpha/2\rceil$ are directed towards the later copy and the rest are undirected~\cite[p.~26503]{deligkas2025}. The undirected edges allow a static walk to \emph{go back in time}. The proof sketch repairs a walk $W_1$ that goes back in time by exchanging suffixes with another walk $W_2$ that ``entered to the left of $v_{t_i}$ and exits to the right''~\cite[p.~26503]{deligkas2025}. The argument assumes $W_2\ne W_1$.

In Fig.~\ref{fig:aaai}(a), the prescribed walk starts at $x$ and ends at $y$. Green solid edges have time $1$ and orange dashed edges time $2$. In panel~(b), numbers give the step order of a static covering walk. The orange dashed step~8 goes back in time.

\begin{proposition}\label{prop:aaai}
The instance of Fig.~\ref{fig:aaai}(a) is infeasible, but the static graph $S$ of the reduction has an exact edge cover with the same number of walks and the same terminals.
\end{proposition}

\begin{proof}
The single walk starts at $x$ and ends at $y$, so it traverses $x\to u\to y$ at time $1$. Afterwards no walk is located at $u$, $v$ or $w$, and the triangle violates~(ii) of Theorem~\ref{thm:char}. In $S$, built as in the proof sketch of Theorem~6.3 of~\cite{deligkas2025} with the terminal $s$ attached to $x$ at time $0$ and $e$ attached to $y$ at time $3$ (Fig.~\ref{fig:aaai}(b)), the gadget of $u$ has $\alpha=\min(2,2)=2$ edges between $u_1$ and $u_2$, one directed edge $(u_1,u_2)$ and one undirected edge. The gadgets of $x$ and $y$ have one directed edge each. The static walk
\[
  s\to x_0\to x_1\to u_1\to u_2\to v_2\to w_2\to u_2\to u_1\to y_1\to y_3\to e
\]
covers every edge of $S$ exactly once, and its step $u_2\to u_1$ (step~8 in Fig.~\ref{fig:aaai}(b)) goes back from time $2$ to time $1$. It is the only walk, so the walk that uses the directed edge $(u_1,u_2)$ is $W_1$ itself, and exchanging suffixes with itself only splits off the triangle as a cycle without a walk.
\end{proof}

\begin{lemma}\label{lem:aaai}
In \FIX{}, the undirected edges of $S$ can be oriented so that every non-terminal node is balanced, with start terminal edges directed outwards and end terminal edges directed inwards, if and only if some orientation of $G$ satisfies condition~(i) of Theorem~\ref{thm:char} together with the end condition $B_{v,q_v}=h_v$.
\end{lemma}

\begin{proof}
Fix an orientation of $G$. It also orients the edges $\{u_t,v_t\}$ of $S$. At a vertex $v$, the start terminal edge contributes $a_v$ incoming edges before the first time label and the end terminal edge contributes $z_v$ outgoing edges after the last time label. Let $n_t$ be the net flow to the right on the chain edges after the copy $v_t$. Balance at $v_t$ gives $n_t=n_{t^-}+\inn_t(v)-\out_t(v)$, where $n_{t^-}=a_v$ at the first copy because the $a_v$ start-terminal edges enter the chain there, so by induction $n_t=c_{t^+}(v)$, and balance at the last copy gives the final counter $z_v$. Between two consecutive time labels there are $\alpha_t=\min(\mathrm{before}_t,\mathrm{after}_t)$ edges with $\mathrm{before}_t=a_v+\delta^G_{\le t}(v)$ and $\mathrm{after}_t=z_v+\delta^G_{>t}(v)$, of which $\lceil\alpha_t/2\rceil$ point to the right and each of the other $\lfloor\alpha_t/2\rfloor$ changes the net flow by $\pm1$. The achievable net flows are exactly the integers $n$ with $n\equiv\alpha_t\pmod 2$ and $\alpha_t\bmod 2\le n\le\alpha_t$. For $n=c_{t^+}(v)$ the parity condition always holds, because $c_{t^+}(v)$ has the parity of both $\mathrm{before}_t$ and $\mathrm{after}_t$ (end condition). The upper bound always holds, because $c_{t^+}(v)$ is at most the number of incoming edges up to $t$ and of outgoing edges after $t$. The lower bound is condition~(i). Conversely, from an orientation that satisfies~(i), choose how many free edges point to the right so that $n_t=c_{t^+}(v)$, which is possible because this value lies in the admissible range.
\end{proof}

Thus, for \FIX{}, balanced orientability of $S$ matches condition~(i) and the end condition (Lemma~\ref{lem:aaai}). In addition $S$ requires static connectivity (every component with edges contains a terminal), which does not depend on the orientation and is weaker than~(ii). A temporal solution yields a cover of $S$, but the converse can fail (Proposition~\ref{prop:aaai}). With the end count dropped (\FREE{}, $a_x=1$, $k=1$), the instance of Fig.~\ref{fig:aaai}(a) is also infeasible, since the single walk must traverse $x\to u\to y$ at time $1$. The static graph $S$ of Proposition~\ref{prop:aaai} attaches the end terminal after the last time label, as in the end-terminal construction shown in the proof sketch and Figure~3 of~\cite{deligkas2025}. The variant with start terminals only is not addressed by this instance, and the solvability claim of Theorem~6.3 of~\cite{deligkas2025} follows by a different argument (Theorem~\ref{thm:reduction}).

\section{M-convex feasibility algorithm}\label{sec:mconvex}

The exchange property of M-convexity lets us reroute pairs of walks between vertices without losing a valid orientation. Submodular flow supplies the standard augmenting path scheme~\cite{cunningham1985,fujishige2005}. Figure~\ref{fig:overview} traces the algorithm on the running example of Section~\ref{sec:ports}.

\subsection{M-convexity}\label{sec:mcsets}

A set $B\subseteq\Z^N$ is \emph{M-convex}~\cite{murota1996,murota2003} if for all $x,y\in B$ and every $i$ with $x_i>y_i$ there is $j$ with $x_j<y_j$ such that $x-\e_i+\e_j\in B$ and $y+\e_i-\e_j\in B$. Here $\e_i$ is the $i$-th unit vector.

\paragraph{Pair counting.}
Write $c_t(v)=2q_t(v)+p_t(v)$ with $p_t(v)\in\{0,1\}$ fixed and $q_t(v)$ the number of \emph{pairs} of walks at $v$. For a component $K$ of $E_t$ and $v\in K$, let $p^{\inn}_v=p_t(v)$, $p^{\out}_v=p_{t^+}(v)$, $\alpha_v=q_t(v)$ (the \emph{incoming} pairs) and $\beta_v=q_{t^+}(v)$ (the \emph{outgoing} pairs). An orientation of $K$ determines the difference $\beta-\alpha$ through
\begin{equation}\label{eq:balance}
  \inn_t(v)-\out_t(v)=2(\beta_v-\alpha_v)+p^{\out}_v-p^{\inn}_v\qquad (v\in K).
\end{equation}
We say that $g\in\Z^K$ is \emph{realizable} if some orientation of $K$ satisfies~\eqref{eq:balance} with $g=\beta-\alpha$.

\begin{definition}\label{def:gamma}
Let $\Gamma_K\subseteq\Z^{K}\times\Z^{K}$ be the set of vectors $(-\alpha,\beta)$ with $\alpha,\beta\ge0$ such that $\beta-\alpha$ is realizable. If $K$ is even, let $\Gamma'_K=\Gamma_K\setminus\{0\}$.
\end{definition}

If $K$ is even, then all walk counters in $K$ are even before and after $t$, so $c_t=2\alpha$ on $K$. Summing~\eqref{eq:balance} over $K$ gives $\beta(K)=\alpha(K)$, where $x(X)=\sum_{i\in X}x_i$. Condition~(ii) at $K$ is therefore exactly $(-\alpha,\beta)\ne0$.

\begin{example}\label{ex:digon}
Let $K$ be two parallel edges between $u$ and $w$, with all parities $0$. Then $\Gamma_K$ contains $(-\e_u,\e_w)$, $(-\e_w,\e_u)$ and $(-\alpha,\alpha)$ for all $\alpha\ge0$, while $\Gamma'_K$ removes only $\alpha=0$, and the first two points still exchange to the non-zero points $(-\e_u,\e_u)$ and $(-\e_w,\e_w)$.
\end{example}

\paragraph{Tools from discrete convexity.}
The following lemmas are standard in orientation theory~\cite{hakimi1965,frank1980,frank2011} and discrete convex analysis~\cite{murota2003,fujishige2005}. Lemmas~\ref{lem:reversal} and~\ref{lem:boundary} are stated with a citation. Lemmas~\ref{lem:exchange} and~\ref{lem:localmax} are proved briefly because they are used in exactly the form below.

\begin{lemma}\label{lem:reversal}
Let $O$ be an orientation of $K$ that realizes $g$ and let $v\ne w$ be vertices of $K$. Then $g+\e_v-\e_w$ is realizable if and only if $O$ has a directed path from $v$ to $w$.
\end{lemma}

This is the classical path-reversal property of orientations with prescribed in-degrees~\cite{hakimi1965,frank1980,frank2011}: reversing a directed path from $v$ to $w$ changes the in-degrees by $+\e_v-\e_w$, and conversely the reversed arcs of any two orientations with these in-degrees contain such a path.

\begin{lemma}\label{lem:boundary}
Let $N$ be a directed network with integer or infinite capacities, a set $P$ of terminal nodes and \emph{fixed} integer supplies at all non-terminal nodes. The set of net inflow vectors at $P$ of integral feasible flows in $N$ is M-convex when it is non-empty. A direct sum of M-convex sets is M-convex.
\end{lemma}

This is the standard fact that the terminal net-inflow vectors of integral flows form an M-convex set~\cite{murota2003,fujishige2005}. For infinite capacities, truncate all capacities at a bound larger than every value of the two flows that give $x,y$. The statement on direct sums follows from the definition, since an exchange in one summand leaves the others unchanged.

\begin{lemma}\label{lem:exchange}
Let $B$ be M-convex, $x\in B$, and let $x+\e_{i_l}-\e_{j_l}\in B$ for $l=1,\dots,r$, where the $i_l$ are pairwise distinct and the $j_l$ are pairwise distinct. If there are no $l<l'$ with $x+\e_{i_l}-\e_{j_{l'}}\in B$, then $x+\sum_l(\e_{i_l}-\e_{j_l})\in B$.
\end{lemma}

\begin{proof}
This is the simultaneous-exchange argument behind shortest augmenting paths for submodular flow~\cite{cunningham1985,fujishige2005,murota2003}, in the form used here. Pairs with $i_l=j_l$ can be dropped, and the hypothesis gives $i_l\ne j_{l'}$ for $l<l'$. We use induction on $r$, the case $r\le1$ being trivial. Let $x'=x+\e_{i_r}-\e_{j_r}\in B$.

First let $l<r$ and apply the exchange axiom to $x+\e_{i_l}-\e_{j_l}$ and $x'$ at the coordinate $i_l$, where the first point is larger. It yields $k\in\{j_l,i_r\}$, where the first point is smaller, with $x'+\e_{i_l}-\e_k\in B$. If $k=i_r\ne j_l$, then $x+\e_{i_l}-\e_{j_r}\in B$, which the hypothesis excludes. Hence $x'+\e_{i_l}-\e_{j_l}\in B$.

Next let $l<l'<r$ and suppose that $u=x'+\e_{i_l}-\e_{j_{l'}}\in B$. Apply the exchange axiom to $u$ and $x$ at the coordinate $i_l$. It yields $k\in\{j_{l'},j_r\}$ with $x+\e_{i_l}-\e_k\in B$, which the hypothesis excludes in both cases. Hence $x'+\e_{i_l}-\e_{j_{l'}}\notin B$.

The pairs $l<r$ therefore satisfy the hypothesis at $x'$, and induction gives the claim.
\end{proof}

\begin{lemma}\label{lem:localmax}
Let $B$ be M-convex, $x\in B$ and $R$ a set of coordinates. If there are no $i\in R$ and $j\notin R$ with $x+\e_i-\e_j\in B$, then $x(R)\ge y(R)$ for every $y\in B$.
\end{lemma}

\begin{proof}
This is the local optimality criterion for maximizing a linear function over an M-convex set~\cite{murota2003}. Suppose some $y\in B$ has $y(R)>x(R)$ and choose such $y$ with $\lVert y-x\rVert_1$ minimal. Some $i\in R$ has $y_i>x_i$. By the exchange axiom there is $j$ with $y_j<x_j$, $y-\e_i+\e_j\in B$ and $x+\e_i-\e_j\in B$. The hypothesis gives $j\in R$, so $y'=y-\e_i+\e_j$ has $y'(R)=y(R)$ and is closer to $x$, a contradiction.
\end{proof}

\begin{figure}[!tb]
\centering
\includegraphics{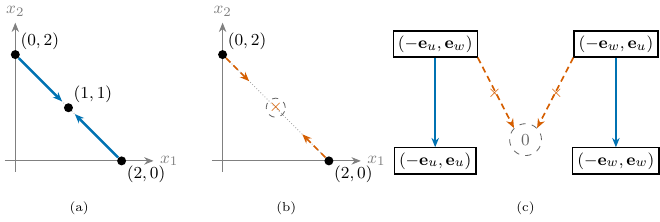}
\caption{Exchange in M-convex sets and removal of the dead state. (a) The set $\{(2,0),(1,1),(0,2)\}$ is M-convex. Its extreme points exchange towards $(1,1)$. (b) After $(1,1)$ is deleted, no exchange remains. (c) For the digon of Example~\ref{ex:digon}, the exchange that would end at the dead state $0$ (dashed) is replaced by an exchange from $(-\e_u,\e_w)$ and $(-\e_w,\e_u)$ to the non-zero states $(-\e_u,\e_u)$ and $(-\e_w,\e_w)$.}
\label{fig:mconvex}
\end{figure}

\paragraph{The main lemma.}
Part~(a) and Lemma~\ref{lem:reversal} are classical~\cite{hakimi1965,frank1980,frank2011}: the in-degree vectors of orientations of a graph form an M-convex set, restated here in pair coordinates. Part~(b) states that deleting the single point $0$ preserves M-convexity when $K$ is connected and Eulerian. Deleting a point from an M-convex set does not preserve M-convexity in general: $\{(2,0),(1,1),(0,2)\}$ is M-convex, but $\{(2,0),(0,2)\}$ is not. To our knowledge, part~(b) is not stated in the standard orientation theory~\cite{frank1980,frank2011}. The submodular flow framework itself is standard~\cite{edmonds1977,franktardos1988,cunningham1985,fujishige2005}. The new ingredient is the handling of the dead state, whose proof also underlies the certificates below. Figures~\ref{fig:mconvex}(a),(b) illustrate the example above, and Figure~\ref{fig:mconvex}(c) illustrates the digon exchange.

For $X\subseteq K$ let $\delta_K(X)$ be the set of edges of $K$ with exactly one end in $X$. The only possible failure after deleting $0$ is that an exchange guaranteed in $\Gamma_K$ makes one of the two states zero.

\begin{lemma}\label{lem:gamma}
\begin{enumerate}
\item[(a)] $\Gamma_K$ is M-convex.
\item[(b)] If $K$ is a (connected) even component, then $\Gamma'_K=\Gamma_K\setminus\{0\}$ is M-convex.
\end{enumerate}
\end{lemma}

\begin{proof}
(a) The statement is classical~\cite{hakimi1965,frank1980,frank2011}. We only reduce it to Lemma~\ref{lem:boundary}. Fix a reference orientation $R$ of $K$ with $d_R=\out-\inn$. Reversing the arc $x\to y$ moves one pair from $y$ to $x$, so every orientation gives $\beta-\alpha=\sigma+\partial\varphi$, where $\sigma(v)=(p^{\inn}_v-p^{\out}_v-d_R(v))/2\in\Z$ and $\varphi\in\{0,1\}$ is the indicator of the reversed arcs of the digraph $R^-$ obtained by reversing every arc of $R$. Build a network on $K$ with the arcs of $R^-$ at capacity $1$, fixed supply $\sigma(v)$ at every $v$, and for each $v$ terminals $v_{\inn}$ and $v_{\out}$ with uncapacitated arcs $v_{\inn}\to v\to v_{\out}$. Its terminal net-inflow vectors are exactly the points $(-\alpha,\beta)$ of $\Gamma_K$, which is M-convex by Lemma~\ref{lem:boundary}.

(b) \emph{Idea.} By~(a) only an exchange that produces the point $0$ is a problem. We show that then another exchange index exists, using reachability in an orientation of $g$ and a strongly connected Euler orientation. All parities in an even component are $0$, so \eqref{eq:balance} reads $\inn_t-\out_t=2g$ with $g=\beta-\alpha$. We use two consequences. First, let $O$ realize $g$ and let $\emptyset\ne X\subsetneq K$. Since $2g(v)=\inn_t(v)-\out_t(v)$, summing over $X$ cancels internal edges and gives $2g(X)=|\{\text{cut edges entering }X\}|-|\{\text{cut edges leaving }X\}|$. Thus if every edge of $\delta_K(X)$ enters $X$, then $g(X)=|\delta_K(X)|/2>0$, and if every such edge leaves, then $g(X)=-|\delta_K(X)|/2<0$. Strictness follows from connectedness of $K$. Second, all degrees in $K$ are even, so $K$ has an Euler circuit~\cite{hierholzer1873}. The resulting orientation realizes $0$ and is strongly connected, so $\e_v-\e_w$ is realizable for all $v,w\in K$ by Lemma~\ref{lem:reversal}.

Let $x,y\in\Gamma'_K$ and let $i$ be a coordinate with $x_i>y_i$. By~(a) there is $j$ with $x_j<y_j$ such that $x'=x-\e_i+\e_j\in\Gamma_K$ and $y'=y+\e_i-\e_j\in\Gamma_K$. If both are non-zero we are done. They cannot both be zero, since then $x=\e_i-\e_j=-y$, while incoming coordinates are non-positive and outgoing coordinates are non-negative, so $i$ would be both an outgoing coordinate ($x_i=1$) and an incoming coordinate ($y_i=-1$). Points of $\Gamma_K$ satisfy $\alpha(K)=\beta(K)$, so a point of $\Gamma'_K$ has $\alpha\ne0$ and $\beta\ne0$.

\emph{Case $x'=0$.} Then $x=(-\e_u,\e_w)$, $i=w_{\out}$ and $j=u_{\inn}$. Write $y=(-\alpha,\beta)$, let $g=\beta-\alpha$ and let $O$ realize $g$. Since $y_i<1$, we have $\beta_w=0$. Let $M$ be the set of vertices reachable from $w$ in $O$. If $M\ne K$, every edge of $\delta_K(M)$ enters $M$, so $\beta(M)\ge g(M)>0$. If $M=K$, then $\beta(K)\ge1$. So there is $z\in M\setminus\{w\}$ with $\beta_z\ge1$, because $\beta_w=0$ and the positive total $\beta(M)$ must occur at another vertex. And $g+\e_w-\e_z$ is realizable by Lemma~\ref{lem:reversal}. Take $j=z_{\out}$ instead, which satisfies $x_j=0<y_j$. Then $x-\e_i+\e_j=(-\e_u,\e_z)\in\Gamma'_K$ by the second consequence, and $y+\e_i-\e_j\in\Gamma'_K$ because its difference is $g+\e_w-\e_z$ and $\alpha\ne0$.

\emph{Case $y'=0$.} Then $y=(-\e_u,\e_w)$, $i=u_{\inn}$ and $j=w_{\out}$. Write $x=(-\alpha,\beta)$, let $g=\beta-\alpha$ and let $O$ realize $g$. Since $x_i>-1$, we have $\alpha_u=0$. Let $\Lambda$ be the set of vertices from which $u$ is reachable in $O$. If $\Lambda\ne K$, every edge of $\delta_K(\Lambda)$ leaves $\Lambda$, so $\alpha(\Lambda)\ge-g(\Lambda)>0$. If $\Lambda=K$, then $\alpha(K)\ge1$. So there is $v\in\Lambda\setminus\{u\}$ with $\alpha_v\ge1$, because $\alpha_u=0$ and the positive total $\alpha(\Lambda)$ must occur at another vertex. And $g+\e_v-\e_u$ is realizable by Lemma~\ref{lem:reversal}. Take $j=v_{\inn}$ instead, which satisfies $x_j=-\alpha_v<0=y_j$. Then $x-\e_i+\e_j$ moves one incoming pair from $v$ to $u$, its difference is $g+\e_v-\e_u$, and it lies in $\Gamma'_K$ (it is non-zero, since its coordinate at $u_{\inn}$ is $-1$). Finally $y+\e_i-\e_j=(-\e_v,\e_w)\in\Gamma'_K$ by the second consequence.
\end{proof}

\begin{remark}\label{rem:connected}
Connectivity is necessary in Lemma~\ref{lem:gamma}(b): for two disjoint triangles treated as one component, the corresponding set is not M-convex. If $g=\beta-\alpha$ is fixed first and the point $0$ is then removed, the exchange axiom fails already for two vertices joined by two parallel edges, so the lemma is stated for $\Gamma'_K$ with $g$ free.
\end{remark}

\subsection{The port network}\label{sec:ports}

Every vertex $v$ becomes a \emph{timeline} of ports. Let $K_1,\dots,K_{q_v}$ be the components that contain the occurrences of $v$. The ports of $v$ are $S_v$ (start), $\IN(K_j,v)$ and $\OUT(K_j,v)$ for $j=1,\dots,q_v$, and $Z_v$ (end). \emph{Waiting arcs} join them in time order, namely $S_v\to\IN(K_1,v)$, $\OUT(K_j,v)\to\IN(K_{j+1},v)$ and $\OUT(K_{q_v},v)\to Z_v$. A vertex without edges has only the arc $S_v\to Z_v$. A flow $\xi\ge0$ on a waiting arc is the number of pairs that wait at $v$ between two occurrences (Fig.~\ref{fig:ports}). For a port $P$ let $\partial\xi(P)$ be the flow out of $P$ minus the flow into $P$, the opposite sign to the net inflow $\partial\varphi$ in the proof of Lemma~\ref{lem:gamma}(a). At the ports of a component, $\partial\xi=(-\alpha,\beta)$.

Let $p_{\mathrm{end}}(v)$ be the parity of $a_v+G_v$. The constraint set $\cB$ is the direct sum of the following sets.
\begin{itemize}
\item At $S_v$, the single point $\floor{a_v/2}$.
\item At the ports of each component $K$, the set $\Gamma'_K$ if $K$ is even and $\Gamma_K$ otherwise.
\item At the ports $Z_v$, for \FIX{}, the single point $-(z_v-p_{\mathrm{end}}(v))/2$ at every $v$, after checking $z_v\equiv a_v+G_v \pmod 2$ and $z_v\ge p_{\mathrm{end}}(v)$. For \FREE{}, the simplex $\{-\gamma:\gamma\in\N^V,\ \gamma(V)=(k-\sum_vp_{\mathrm{end}}(v))/2\}$, which fixes only the total number of final pairs.
\end{itemize}
The single points and the end simplex of \FREE{}, the integer points of the uniform base polytope $\{-\gamma:\gamma\ge0,\ \gamma(V)=c\}$, are M-convex~\cite{murota2003}. By Lemmas~\ref{lem:boundary} and~\ref{lem:gamma}, $\cB$, their direct sum, is M-convex. It is non-empty after the checks of line~1 of Algorithm~\ref{alg:pf}: arbitrary orientations give states in each $\Gamma_K$, and if an even component would have the zero state, adding one waiting pair at the same vertex on both sides gives a nonzero state in $\Gamma'_K$. The start and end sets are nonempty by the parity and total-count checks. Since $\sum_v(a_v+G_v)=k+2m$, the number $k-\sum_vp_{\mathrm{end}}(v)$ is always even.

\begin{figure}[!tb]
\centering
\includegraphics{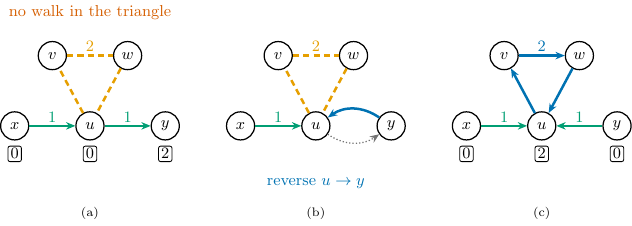}
\caption{Overview of the algorithm on the running example. The boxes show the number of walks at each vertex after time 1: $x:0$, $u:0$, $y:2$ in (a) and $x:0$, $u:2$, $y:0$ in (c). (a) After the maximum flow, the time-1 edges leave no walk at $u$, so the time-2 triangle is dead. (b) An augmenting path reverses the orientation of $u\to y$. (c) After augmentation, $u$ holds two walks: $x\to u\to v\to w\to u$ and $y\to u$.}
\label{fig:overview}
\end{figure}

\begin{figure}[!htbp]
\centering
\includegraphics{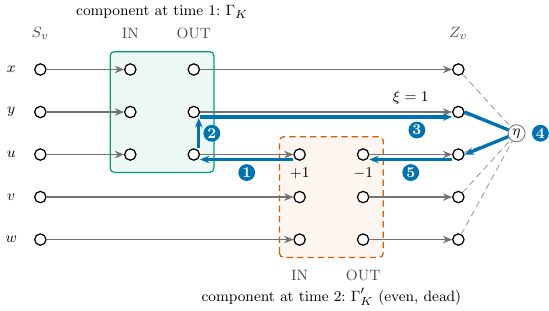}
\caption{Port network and an augmenting path for the running \FREE{} example.}
\label{fig:ports}
\end{figure}

Running example. Take Fig.~\ref{fig:aaai}(a) with one walk starting at each of $x$ and $y$ and free ends. Figure~\ref{fig:overview} traces its three stages. Figure~\ref{fig:ports} shows the port network for this \FREE{} instance and an augmenting path (thick blue arrows $1$--$5$) that repairs its dead triangle; $+1$ and $-1$ mark nonzero deviations, and $\eta$ implements an exchange between end ports.

\begin{theorem}\label{thm:reduction}
An instance of \FREE{} or \FIX{} is feasible if and only if there is a pair flow $\xi\ge0$ on the waiting arcs with $\partial\xi\in\cB$.
\end{theorem}

\begin{proof}
($\Rightarrow$) Take the orientation of the walks and let $\xi=q$ on every waiting arc. At $S_v$, $\partial\xi=\floor{a_v/2}$. At each component, $\partial\xi=(-\alpha,\beta)$ with $\beta\ge0$ by~(i), the difference $\beta-\alpha$ is realized by the orientation itself, and $(-\alpha,\beta)\ne0$ on even components by~(ii). At $Z_v$, the value $-\partial\xi$ is the number of final pairs at $v$. It equals $(z_v-p_{\mathrm{end}}(v))/2$ in \FIX{}, and in \FREE{} these values sum to $(k-\sum_vp_{\mathrm{end}}(v))/2$ because the number of walks is preserved.

($\Leftarrow$) For each $K$ choose an orientation that realizes $\beta-\alpha$. By~\eqref{eq:balance}, the number of walks at $v$ between two occurrences under the combined orientation equals $2\xi+p$ on the corresponding waiting arc. Condition~(i) holds because $\beta\ge0$. Condition~(ii) holds on components that are not even by Lemma~\ref{lem:noneven} and on even components because $(-\alpha,\beta)\ne0$. In \FIX{} the final number of walks at $v$ is $z_v$. Theorem~\ref{thm:char} completes the proof.
\end{proof}

Equivalently, Theorem~\ref{thm:reduction} asks whether the M-convex set $\{\partial\xi:\xi\ge0\}$ of boundaries of non-negative waiting flows (M-convex by Lemma~\ref{lem:boundary}, up to sign) meets the M-convex set $\cB$. This is a feasibility instance of M-convex intersection~\cite{murota2003,edmonds1977,franktardos1988}.

\subsection{The augmenting path algorithm}\label{sec:auxgraph}

\begin{figure}[!tb]
\centering
\includegraphics{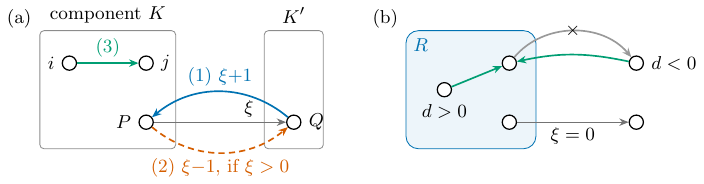}
\caption{The auxiliary graph. (a) The three kinds of arcs at a waiting arc $P\to Q$ with flow $\xi$ and at two ports $i,j$ of one component, where kind~(3) requires $x+\e_i-\e_j\in\cB$; each arc moves one unit of deviation from its tail to its head. (b) The certificate of Theorem~\ref{thm:correct}: the set $R$ of ports reachable from $\{d>0\}$ contains no port with $d<0$, and no arc leaves $R$. Then $x(R)=-d(R)<0$, whereas Lemma~\ref{lem:localmax} would give $x(R)\ge x'(R)\ge0$ for any solution $(\xi',x')$.}
\label{fig:aux}
\end{figure}

The algorithm initializes $\xi$ from the chain-flow orientation: on each waiting arc it puts the number of pairs (half the walk counter, an integer by the fixed parity). The IN and OUT coordinates of each component in $x$ are $-\alpha$ and $\beta$ from that orientation, $x(S_v)=\lfloor a_v/2\rfloor$, and the end coordinate is minus the final pair count. For each initially dead even component, line 3 replaces its zero state by one incoming and outgoing pair at a chosen vertex. The algorithm maintains a flow $\xi\ge0$ and a point $x\in\cB$. Their \emph{deviation} is $d=\partial\xi-x$. A solution is obtained when $d=0$. The \emph{auxiliary graph} has the ports as nodes and three kinds of arcs, each of which moves one unit of deviation from its tail to its head.
\begin{enumerate}
\item[(1)] The reverse of every waiting arc. Using it increases $\xi$ on that arc by one.
\item[(2)] Every waiting arc with $\xi>0$. Using it decreases $\xi$ by one.
\item[(3)] An exchange arc $i\to j$ whenever $x+\e_i-\e_j\in\cB$. Using it replaces $x$ by $x+\e_i-\e_j$.
\end{enumerate}
Let $i$ be a port of $v$ and $j$ a port of $w$ in the same component $K$. The exchange arc $i\to j$ exists if and only if $\alpha_v\ge1$ when $i$ is an incoming port, $\beta_w\ge1$ when $j$ is an outgoing port, $v=w$ or the current orientation of $K$ has a directed path from $v$ to $w$ (Lemma~\ref{lem:reversal}), and, if $K$ is even, the new point is non-zero. No availability condition is needed when $i$ is an outgoing port or $j$ is an incoming port, since those changes increase the corresponding pair count. In the simplex of \FREE{}, $x+\e_i-\e_j\in\cB$ whenever $\gamma_i\ge1$. Instead of these $\Theta(n^2)$ arcs the auxiliary graph has a node $\eta$ with arcs $Z_i\to\eta$ for every $i$ with $\gamma_i\ge1$ and $\eta\to Z_j$ for every $j$, and a path through $\eta$ performs one simplex exchange. Any shortest path and any vertex selected in line 3 may be used. Figure~\ref{fig:aux} illustrates the three kinds of arcs and the certificate used in Theorem~\ref{thm:correct}.

\begin{algorithm}[t]
\caption{Augmenting path algorithm for \FIX{} and \FREE{}}\label{alg:pf}
\begin{algorithmic}[1]
\State Check the counting conditions ($\sum a=\sum z$ and $h_v\in\N$ for \FIX{}, $k\ge\sum_vp_{\mathrm{end}}(v)$ for \FREE{}). \textbf{if} a check fails \textbf{then return} \NO{}.
\State Solve a maximum flow in the chain network. \textbf{if} its value is less than $m$ \textbf{then return} \NO{} with the minimum cut.
\State Take the orientation of the maximum flow and the induced $\xi$ and $x$. For every dead even component $K$, pick $v\in K$, set $x=(-\e_v,\e_v)$ at $(\IN(K,v),\OUT(K,v))$, so that $d=+1$ at $\IN(K,v)$ and $d=-1$ at $\OUT(K,v)$.
\While{some port has $d>0$}
  \State Find a \emph{shortest} path from $\{d>0\}$ to $\{d<0\}$ in the auxiliary graph by breadth-first search.
  \If{no such path exists}
    \State \Return \NO{} together with the set $R$ of ports reachable from $\{d>0\}$.
  \EndIf
  \State Apply all arcs of the path to $\xi$ and $x$ at once and re-orient the components whose point changed (see below).
\EndWhile
\State \Return \YES{} and build the walks as in the proof of Theorem~\ref{thm:char}.
\end{algorithmic}
\end{algorithm}

In the theorem-bound re-orientation, one repeatedly reverses a directed path from a vertex whose target in-degree exceeds its current in-degree to one whose current in-degree exceeds its target, which reduces the $L_1$ distance to the target by two; $O(|K|)$ paths of $O(m_K)$ time each suffice for a component $K$.

\subsection{Correctness and infeasibility certificates}

\begin{lemma}\label{lem:invariant}
After every augmentation, $\xi\ge0$ and $x\in\cB$.
\end{lemma}

\begin{proof}
All arcs of the path are evaluated before the augmentation and applied together. A shortest path is simple, so every waiting arc is used at most once and arcs of kind~(2) are used only when $\xi>0$. Hence $\xi\ge0$. A simple path visits $\eta$ at most once, so the simplex of \FREE{} undergoes at most one exchange, which is valid by the definition of the arcs. Consider any other summand $\Gamma$ of $\cB$ and its exchanges $(i_1,j_1),\dots,(i_r,j_r)$ in the order of the path. The nodes of a shortest path have pairwise distinct and increasing breadth-first distances from the set $\{d>0\}$, so the $i_l$ are pairwise distinct and the $j_l$ are pairwise distinct. If there were $l<l'$ with an arc $i_l\to j_{l'}$, then $\mathrm{dist}(j_{l'})\le\mathrm{dist}(i_l)+1<\mathrm{dist}(i_{l'})+1=\mathrm{dist}(j_{l'})$, a contradiction. So the hypothesis of Lemma~\ref{lem:exchange} holds. The argument uses only the order of distances, so it is unaffected when the path enters and leaves a component several times, and the lemma is applied to $\Gamma'_K$ itself, which is M-convex by Lemma~\ref{lem:gamma}. By Lemma~\ref{lem:exchange}, the new point stays in the set.
\end{proof}

\begin{theorem}\label{thm:correct}
Algorithm~\ref{alg:pf} answers correctly. When it answers \YES{}, the returned walks form a solution. When it answers \NO{} in the loop, the set $R$ of ports reachable from $\{d>0\}$ certifies infeasibility.
\end{theorem}

\begin{proof}
A \NO{} in line~2 is correct because~(i) is necessary. The sum of $d$ over all ports is $0$ after line~3 (the only non-zero entries are the $+1$ and $-1$ of the repaired dead components) and stays $0$, since every arc of the auxiliary graph moves one unit between two ports, so if the loop ends then $d=0$, $\partial\xi=x\in\cB$ with $\xi\ge0$ by Lemma~\ref{lem:invariant}, and the walks are built through Theorem~\ref{thm:reduction}. Suppose the algorithm stops with the set $R$. It contains every port with $d>0$ and no port with $d<0$, so $d(R)>0$. No arc of the auxiliary graph leaves $R$. By kind~(1), no waiting arc enters $R$, and by kind~(2), every waiting arc that leaves $R$ has $\xi=0$. Hence $\partial\xi(R)=0$ and $x(R)=-d(R)<0$. Since $x\in\cB$ and $\xi\ge0$ throughout (Lemma~\ref{lem:invariant}), assume a solution $(\xi',x')$ exists with $\xi'\ge0$ and $x'=\partial\xi'\in\cB$. Since no waiting arc enters $R$, $x'(R)$ is the total of $\xi'$ on the waiting arcs that leave $R$, so $x'(R)\ge0$. No exchange arc leaves $R$. For the simplex this holds because a port $Z_i\in R$ with $\gamma_i\ge1$ puts $\eta$ and every $Z_j$ into $R$. Lemma~\ref{lem:localmax}, applied to each summand of the direct sum, gives $x(R)\ge x'(R)\ge0$, a contradiction.
\end{proof}

For every \YES{} answer, the implementation constructs the walks and checks them against the problem definition. A \NO{} answer from line~2 has a chain-network cut certificate verifiable from the instance data alone. For a loop \NO{}, a checker verifies $\xi,x$, computes $d=\partial\xi-x$, checks $d(R)>0$ and confirms that no valid waiting or exchange arc leaves $R$. This does not depend on Lemma~\ref{lem:exchange}, since $x\in\cB$ is checked at run time, but it depends on Lemmas~\ref{lem:gamma} and~\ref{lem:localmax}, so its verification is theory dependent.

\subsection{Complexity}

\begin{theorem}\label{thm:time}
Let $D$ be the number of dead even components after the maximum flow and $s$ the largest number of vertices of a snapshot component. Algorithm~\ref{alg:pf} performs at most $D$ augmentations and uses
\[
  O\bigl(n+m\log(m+1)+m^2+D\,(n+s\,m)\bigr)\subseteq O(n+n\,m^2)
\]
arithmetic operations, since $D\le m/2$ and $s\le n$.
\end{theorem}

\begin{proof}
Each augmentation decreases the sum of positive deviations by exactly one, and that sum is $D$ after line~3. Distinct dead even components are edge-disjoint and each has at least two edges, so $D\le m/2$. Lines 1 to 3 take $O(n+m^2)$ arithmetic operations, dominated by the initial maximum flow: the chain network has $O(m)$ nodes and arcs and the flow value is at most $m$, so at most $m$ augmenting paths of $O(m)$ time each suffice. One augmentation costs $O(n+sm)$. The number of ports is $O(n+m)$, so the breadth-first search on the auxiliary graph takes $O(n+m)$ time apart from the exchange tests. For a component $K$ with $m_K$ edges, depth-first search from every vertex of $K$ takes $O(\abs{K}\,m_K)$ time, and the $O(\abs{K}^2)$ candidate exchange pairs cost the same, since $\abs{K}\le m_K+1$. Re-orienting a changed component $K$ needs $O(\abs{K})$ path reversals of $O(m_K)$ time each, because the exchanges inside $K$ use distinct ports and $K$ has $2\abs{K}$ ports: reversing a path from a surplus to a deficit lowers the $L_1$ distance to the target boundary by two (Lemma~\ref{lem:reversal}). Summing over components, $\sum_K\abs{K}\,m_K\le sm$.
\end{proof}

The bound does not depend on the magnitudes of $a$ and $z$: every number handled is an integer of absolute value $O(k+m)$, so it is also polynomial in the bit size of the input. When $D=0$ the algorithm is a single maximum flow. The value of $D$ depends on the initial maximum flow, since the orientation of any solution is itself a maximum flow with $D=0$.

\section{Strict walks are NP-complete}\label{sec:strict}

A \emph{strict} walk has strictly increasing time labels and uses at most one edge of each snapshot. Let \SFIX{} and \SFREE{} denote the strict variants of \FIX{} and \FREE{}. The case with prescribed ends was left open by Deligkas et al.~\cite[p.~26504]{deligkas2025}, beyond the special cases of Section~\ref{sec:related}.

\subsection{Realizable orientations}

A strict walk that arrives at a vertex at time $t$ cannot leave it at time $t$.

\begin{lemma}\label{lem:strict}
An orientation is realizable by strict walks that start according to $a$ if and only if $\out_t(v)\le c_t(v)$ for every vertex $v$ and every time $t$. For \SFIX{} the final number of walks at $v$ must in addition equal $z_v$ for every $v$.
\end{lemma}

\begin{proof}
\emph{Necessity.} A strict walk uses at most one edge at time $t$. The $\out_t(v)$ arcs that leave $v$ at time $t$ are therefore traversed by pairwise distinct walks, and each of these walks is located at $v$ just before $t$. \emph{Sufficiency.} Process the times in increasing order. At time $t$, assign to the arcs that leave $v$ pairwise distinct walks located at $v$, which exist because $\out_t(v)\le c_t(v)$. Each of these walks traverses its arc and all other walks wait. The number of walks at every vertex after $t$ equals $c_{t^+}$, so the induction continues.
\end{proof}

Compared with Theorem~\ref{thm:char}, condition~(i) is replaced by the stronger condition $\out_t(v)\le c_t(v)$, and condition~(ii) disappears, since every edge is traversed by a walk that was at its tail before. Thus an orientation satisfying the strict condition has no dead even component. In the notation of Section~\ref{sec:char} the new condition reads $2B_{v,j-1}+b_{v,j}\le a_v+G_{v,j-1}$. The different weights on current and earlier departures prevent expressing it as a capacity constraint of the chain network.

\subsection{The reduction}

\begin{figure}[!htbp]
\centering
\includegraphics{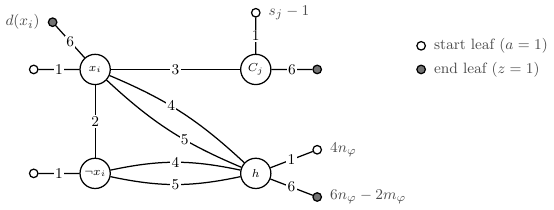}
\caption{The strict-walk reduction around a variable and an incident clause.}
\label{fig:strict}
\end{figure}

We reduce from satisfiability with clauses of two or three literals in which every variable occurs at most three times, which is NP-complete by Tovey~\cite{tovey1984}. We may assume that the literals of a clause belong to distinct variables (discard tautological clauses, merge repeated literals, propagate unit clauses), and that every literal occurs at most twice: a literal that occurs three times is pure, because its variable occurs at most three times and so the opposite literal does not occur, hence we set it to true and delete its clauses. This creates no new occurrences and no clause of size $1$. Let $\varphi$ be such a formula with variables $x_1,\dots,x_{n_\varphi}$ and clauses $C_1,\dots,C_{m_\varphi}$. Let $s_j\in\{2,3\}$ be the number of literals of $C_j$ and let $d(\ell)\le2$ be the number of clauses that contain the literal $\ell$. Since $\sum_js_j=\sum_\ell d(\ell)\le4n_\varphi$ and $s_j\ge2$, we have $m_\varphi\le2n_\varphi$. The temporal graph $G_\varphi$ has one vertex for each of the $2n_\varphi$ literals, one vertex for each clause, a \emph{hub} $h$, and vertices of degree one called \emph{leaves}. Its edges are the following (Fig.~\ref{fig:strict}).
\begin{itemize}
\item \emph{Time $1$.} One \emph{start leaf} is joined to every literal vertex, $s_j-1$ start leaves are joined to $C_j$, and $4n_\varphi$ start leaves are joined to $h$.
\item \emph{Time $2$.} The edge $\{x_i,\lnot x_i\}$ for every variable $x_i$.
\item \emph{Time $3$.} The edge $\{C_j,\ell\}$ for every clause $C_j$ and every literal $\ell$ of $C_j$.
\item \emph{Times $4$ and $5$.} One edge $\{\ell,h\}$ at time $4$ and one at time $5$, for every literal vertex $\ell$.
\item \emph{Time $6$.} $d(\ell)$ \emph{end leaves} are joined to every literal vertex $\ell$, one end leaf is joined to every $C_j$, and $6n_\varphi-2m_\varphi$ end leaves are joined to $h$.
\end{itemize}
We set $a_v=1$ at every start leaf, $z_v=1$ at every end leaf, and $a_v=z_v=0$ otherwise. There are $6n_\varphi+\sum_js_j-m_\varphi$ start leaves and as many end leaves. The graph $G_\varphi$ has no parallel edges within a snapshot, and its size is linear in the size of $\varphi$.

In Fig.~\ref{fig:strict}, edge labels are times and numbers beside leaves are multiplicities. White leaves are start leaves and shaded leaves are end leaves. Other clause incidences and the end leaves of $\lnot x_i$ are omitted.

\begin{theorem}\label{thm:strict}
\SFIX{} is NP-complete. It remains NP-complete when the temporal graph has six time labels and no parallel edges within a snapshot, and when every vertex with $a_v+z_v>0$ is a leaf with $a_v+z_v=1$, so that every walk of every solution is non-empty. \SFREE{} is NP-complete even for temporal graphs with two time labels, and with three time labels if empty walks are not allowed.
\end{theorem}

\begin{proof}
Both problems are in NP: an orientation is a certificate checked in polynomial time by Lemma~\ref{lem:strict}, and when empty walks are not allowed the certificate also lists the at most $m$ non-empty walks. We show that $\varphi$ is satisfiable if and only if $G_\varphi$ with the vectors $a$ and $z$ above is a feasible instance of \SFIX{}.

\emph{A solution yields a satisfying assignment.} Take the orientation of a solution. Before time $1$ only the start leaves hold walks, so the condition $\out_1(v)\le c_1(v)$ of Lemma~\ref{lem:strict} forces every edge at time $1$ to be directed away from its leaf, since its other end holds no walk. Before time $2$ every literal vertex therefore holds one walk, $C_j$ holds $s_j-1$ walks and $h$ holds $4n_\varphi$ walks. The edge $\{x_i,\lnot x_i\}$ at time $2$ moves one walk, so before time $3$ one of the two vertices $x_i$ and $\lnot x_i$ holds two walks and the other holds none. Call a literal \emph{true} if its vertex holds two walks. Exactly one of $x_i$ and $\lnot x_i$ is true, so this is a truth assignment. At time $3$ the vertex $C_j$ has $s_j$ edges but holds only $s_j-1$ walks. By Lemma~\ref{lem:strict} at least one of these edges is directed from a literal vertex $\ell$ to $C_j$, and $\ell$ holds a walk before time $3$. Hence $\ell$ is true and $C_j$ is satisfied.

\emph{A satisfying assignment yields a solution.} We give an orientation and verify Lemma~\ref{lem:strict}. Direct the edges at time $1$ away from the leaves and the edge $\{x_i,\lnot x_i\}$ from the false literal to the true literal. For every clause $C_j$ choose one true literal $\ell_j$ of $C_j$, direct $\{C_j,\ell_j\}$ from $\ell_j$ to $C_j$, and direct the other $s_j-1$ edges of $C_j$ away from $C_j$. Let $\theta_\ell\le d(\ell)\le2$ be the number of clauses that choose a true literal $\ell$. A true literal holds two walks, which suffice for its at most two outgoing edges at time $3$, and a false literal has none. After time $3$ the vertex $C_j$ holds one walk, a false literal $\ell$ holds $d(\ell)$ walks, a true literal $\ell$ holds $d(\ell)+2-2\theta_\ell$ walks, and $h$ holds $4n_\varphi$ walks. The edges at times $4$ and $5$ bring every literal vertex $\ell$ to exactly $d(\ell)$ walks.
\begin{itemize}
\item If $\ell$ is true and $\theta_\ell=0$, direct both edges from $\ell$ to $h$. The vertex $\ell$ holds $d(\ell)+2\ge2$ walks before time $4$.
\item If $\ell$ is false, or true with $\theta_\ell=1$, direct the edge at time $4$ from $h$ to $\ell$ and the edge at time $5$ from $\ell$ to $h$. The vertex $\ell$ holds at least one walk before time $5$.
\item If $\ell$ is true and $\theta_\ell=2$, direct both edges from $h$ to $\ell$.
\end{itemize}
At each of the times $4$ and $5$ at most $2n_\varphi$ edges leave $h$, and $h$ holds $4n_\varphi$ walks before time $4$ and at least $2n_\varphi$ walks before time $5$. The number of walks at $h$ changes in total by $\sum_{\ell\text{ true}}(2-2\theta_\ell)=2n_\varphi-2m_\varphi$, because there are $n_\varphi$ true literals and $\sum_\ell\theta_\ell=m_\varphi$. Hence $h$ holds $6n_\varphi-2m_\varphi\ge0$ walks before time $6$ (using $m_\varphi\le2n_\varphi$). Before time $6$ every vertex that is not a leaf thus holds exactly as many walks as it has end leaves, and the edges at time $6$ are directed towards the end leaves. The final numbers of walks are $z$.

\emph{Free end vertices.} Keep only the literal and clause vertices and the edges of times $2$ and $3$, and set $a_v=1$ at every literal vertex and $a_{C_j}=s_j-1$. Each vertex then holds the same number of walks before time $2$ as before, so the arguments for times $2$ and $3$ apply unchanged and this instance of \SFREE{} is feasible if and only if $\varphi$ is satisfiable. In a solution, a true literal that no clause chooses keeps an empty walk. If empty walks are not allowed, keep in addition the start leaves and the edges of time~$1$, with $a_v=1$ at every start leaf and $a_v=0$ elsewhere. Every walk then starts with an edge, and the same arguments apply with three time labels.
\end{proof}

The graph $G_\varphi$ is not proper and its snapshots have different numbers of edges, so neither Corollary~6.4 nor Theorem~6.5 of~\cite{deligkas2025} applies. The latter is exponential in $k$, which grows with $n_\varphi$ here. Theorem~5.4 of~\cite{deligkas2025} concerns the problem without prescribed start vertices and does not imply Theorem~\ref{thm:strict}. Whether \SFIX{} remains NP-complete when every vertex has at most two edges in every snapshot is left open.

\section{Computational experiments}\label{sec:eval}

We compare pair flow (\textbf{PF}, Algorithm~\ref{alg:pf}) with exhaustive oracles, an integer program (\textbf{ILP}), restart branch and bound (\textbf{BB}; it adds two walks at a dead even component and reruns maximum flow) and the reconstructed static rule of Section~\ref{sec:counterexample}.

\subsection{Validation}\label{sec:exp-correct}

PF \YES{} walks were checked against the definition. Python used scratch re-orientation (default, bound $O(D(n+m^2))$) or reversed directed paths (Theorem~\ref{thm:time}); the scaling runs and paired comparisons use both. BB limits were $10$~s or $2{,}000$ maximum flows per even-rich run ($5$~s at nominal $m=1{,}000$), $200$ maximum flows on census windows, $150$~s on Turku and $60$~s in the ILP comparison. The exhaustive oracle searches snapshot-edge traversals token by token, using neither orientations nor Theorem~\ref{thm:char}, so it tests the characterization against the operational definition. The ILP encodes Theorem~\ref{thm:char} directly, with binary orientation variables and non-negative counters. On the even-rich and shortage families every \NO{} came from augmentation, so its certificate rests on Lemma~\ref{lem:gamma}(b) and Lemma~\ref{lem:localmax} and is not verifiable from instance data.

\begin{table}[!t]
\caption{Validation checks.}
\label{tab:validation}
\footnotesize
\begin{tabular*}{\textwidth}{@{\extracolsep\fill}p{0.25\textwidth}p{0.22\textwidth}p{0.40\textwidth}}
\toprule
Check & Instances/runs & Outcome \\
\midrule
Exhaustive oracle & $102{,}476$ instances; $204{,}952$ runs & All agreed with the operational definition \\
PF--BB agreement & $105{,}182$ runs & Answers agreed \\
Matched comparisons & $10{,}813$ pairs; $4{,}000$ random comparisons & Answers agreed \\
Even-rich suite & $1{,}176$ oracle runs; $2{,}918$ completed BB runs ($42$ stopped) & Answers agreed \\
Python--C++ & $8{,}000$ instances & Answers agreed \\
Strict reduction & $1{,}967$ formulas & Feasibility matched satisfiability \\
Lemma~\ref{lem:gamma} enumeration & $119{,}040$ labelled sets for~(a), $153$ for~(b) & No violation found for connected multigraphs with $n\le5$, $m\le6$ \\
Certificate re-check & $13{,}028$ certificates & $78.85\%$--$99.27\%$ of \NO{} certificates verifiable from instance data in the three checked sets \\
\bottomrule
\end{tabular*}
\end{table}

\paragraph{Limits of the validation.} The oracle and ILP were written by the authors with AI assistance (see Declarations). The ILP and PF share Theorem~\ref{thm:char}, so their agreement tests internal consistency, not independent validation. The Lemma~\ref{lem:gamma} enumeration covers $4{,}065$ labelled graphs for part~(a) but only the $153$ even ones, one set each, for part~(b); it is only a small-box sanity check.

\subsection{Benchmarks and scaling}\label{sec:scope}

The even-rich benchmark isolates augmentation. A graph with nominal edge count $m$ has $\max(1,\mathrm{round}(\sqrt m/2))$ \emph{pairs} of walks, each pair starting together at a random vertex, and $\max(8,\lfloor m/3\rfloor)$ (sparse) or $\max(8,\lfloor 2\sqrt m\rfloor)$ (dense) vertices. At each time $t=1,2,\dots$, until the graph has at least $m$ edges, each pair is skipped with probability $0.3$ or adds at its current vertex $u$ a double edge to a random vertex $v$ (probability $0.25$), a cycle of length $3$--$6$ through $u$ ($0.30$), or two internally disjoint $u$--$v$ paths of one to three edges each ($0.45$); the pair then moves to $v$, except after a cycle. Degrees, walk counters and components are therefore even in every snapshot, so counting alone accepts every \FREE{} instance. Labels and edge order are shuffled, and the end count is $2$ at each pair's final vertex. Each graph (seed $20261113$) yields planted \FIX{} and \FREE{} instances and versions with one start pair removed (and, for \FIX{}, the corresponding end pair): $3{,}056$ instances on $764$ graphs with $12\le m\le20{,}000$ (Table~\ref{tab:even}), run with Python~3.13.12 on an Intel Core i9-9980HK under Linux. Initial flow left dead even components in $1{,}349$ instances ($D$ up to $3{,}310$), and every \NO{} answer came from augmentation rather than counting. On the $1{,}480$ \FIX{} instances in the first eight rows of Table~\ref{tab:even}, PF accepted $1{,}229$ and rejected $251$; our reconstruction of the static rule wrongly accepted $250$ and rejected none that PF accepted.

\begin{table}[!t]
\caption{Even-rich benchmark (Python, default backend without SciPy, from scratch). $m$: nominal edge count. Even: mean number of even snapshot components over all times. Dead: mean/maximum number of dead even components after the initial flow. Cut: pair-removed instances answered \YES{}/\NO{} by PF. PF: median/maximum seconds. BB: stopped/all runs. Static: false \YES{} answers of the reconstructed static rule among PF-\NO{} \FIX{} instances. Dashes: not run.}
\label{tab:even}
\scriptsize
\setlength{\tabcolsep}{2.5pt}
\begin{tabular*}{\textwidth}{@{\extracolsep\fill}lrrrrrrrr}
\toprule
Family & $m$ & Graphs & Even & Dead & Cut & PF & BB & Static \\
\midrule
sparse & $12$ & $200$ & $3.0$ & $0.3$/$4$ & $265$/$135$ & $<0.001$/$<0.001$ & $0$/$800$ & $77$/$78$ \\
dense & $12$ & $200$ & $2.7$ & $0.3$/$4$ & $280$/$120$ & $<0.001$/$<0.001$ & $0$/$800$ & $72$/$72$ \\
sparse & $50$ & $100$ & $7.4$ & $0.7$/$5$ & $165$/$35$ & $<0.001$/$0.002$ & $0$/$400$ & $23$/$23$ \\
dense & $50$ & $100$ & $7.1$ & $0.6$/$4$ & $172$/$28$ & $<0.001$/$0.002$ & $0$/$400$ & $19$/$19$ \\
sparse & $200$ & $50$ & $35.9$ & $10.0$/$24$ & $50$/$50$ & $0.007$/$0.014$ & $10$/$200$ & $28$/$28$ \\
dense & $200$ & $50$ & $21.9$ & $3.4$/$12$ & $83$/$17$ & $0.004$/$0.011$ & $2$/$200$ & $11$/$11$ \\
sparse & $1{,}000$ & $20$ & $218.8$ & $90.8$/$117$ & $3$/$37$ & $0.157$/$0.215$ & $30$/$80$ & $20$/$20$ \\
dense & $1{,}000$ & $20$ & $78.4$ & $16.1$/$27$ & $40$/$0$ & $0.079$/$0.132$ & $0$/$80$ & $0$/$0$ \\
sparse & $5{,}000$ & $10$ & $1242.1$ & $682.3$/$733$ & $0$/$20$ & $4.81$/$7.29$ & -- & -- \\
dense & $5{,}000$ & $10$ & $305.1$ & $77.5$/$94$ & $19$/$1$ & $1.74$/$2.20$ & -- & -- \\
sparse & $20{,}000$ & $2$ & $5183.5$ & $3277.6$/$3{,}310$ & $0$/$4$ & $154.8$/$170.5$ & -- & -- \\
dense & $20{,}000$ & $2$ & $1098.5$ & $309.5$/$315$ & $2$/$2$ & $51.3$/$56.8$ & -- & -- \\
\bottomrule
\end{tabular*}
\end{table}

\paragraph{Scaling of the two re-orientations.}
In the shortage family, a hub with $2(e-1)$ walks must serve $e$ gadgets that need $2e$, so every \FREE{} instance is \NO{}; here $m=7e$, component size is constant and $D$ grows proportionally to $m$ ($62$ to $499$ augmentations for $e=250$ to $2{,}000$). With the SciPy-assisted backend on a shared machine and two repetitions per size, scratch times grew from $0.84$--$1.08$~s at $m=1{,}750$ to $98.0$--$99.9$~s at $m=14{,}000$, and incremental times from $0.74$--$0.85$~s to $64.6$--$66.1$~s. Log-log slopes between consecutive sizes within a repetition were $1.81$--$2.84$, and $2.08$--$2.30$ from $e=250$ to $e=2{,}000$, consistent with the near-$m^2$ growth predicted by $O(D(n+sm))$; two repetitions do not verify Theorem~\ref{thm:time}. The incremental variant was faster in all eight paired repetitions ($1.04$ to $1.5$ times), and the mean ratio on paired even-rich instances was $1.03$--$1.17$.

\paragraph{Integer programming baseline.} The ILP, solved by HiGHS~\cite{huangfu2018} (SciPy~1.16.3), agreed with PF and the oracle on $3{,}000$ random instances (at most $10$ edges, $484$ with an even component) and with PF on $82$ benchmark cases ($10$ shortage and $72$ even-rich), where BB stopped at its $60$~s limit in eight. With the SciPy-assisted backend on a shared machine (unlike Table~\ref{tab:even}), median ILP/PF/BB times on even-rich graphs with $m=1{,}000$ were $0.110/0.252/0.610$~s (sparse) and $0.373/0.154/0.220$~s (dense); the ILP was not run above $m=1{,}400$.

\subsection{Real data}

Contact data~\cite{sapiezynski2019,stehle2011,fournet2014,mastrandrea2015,vanhems2013} were treated as undirected, with duplicate and self-contacts removed (Copenhagen Bluetooth: $5$-minute bins, RSSI $\ge-75$). Start vectors are planted: the \emph{greedy} scheme scans snapshots in time order, and an endpoint walk takes each edge or is created if none is there; the \emph{tight} variants drop the end vector and remove one walk or ten percent of the starts to exercise \NO{} answers. In transit timetables~\cite{kujala2018}, each vehicle trip is a walk from first to last stop, with one edge per segment at departure time. We used $304$ one-hour windows (four cities, four schemes), $12$ whole-day instances (three cities) and $50$ larger \FIX{} and \FREE{} instances on $25$ cities, using the busiest three-hour block for the $12$ cities above $250{,}000$ daily trip segments.

Dead even components occurred in $48$ of the $50$ larger and $6$ of the $12$ whole-day transit instances, and rarely elsewhere. Every \NO{} answer on real data came from counting maximum flow, so these runs test scalability, not the M-convex machinery. On the whole-day Turku timetable ($125{,}779$ edges, Python), PF solved \FIX{} in $70.3$ and \FREE{} in $33.7$ seconds, while BB did not finish within $150$ seconds. C++17 solved all $50$ larger instances; the largest, Paris \FIX{} with $387{,}467$ edges and $D=287$, took $68.7$ seconds in a single run. These runs used an Intel Core i5-10210U (four cores, Linux under WSL2) with two to four concurrent jobs, so times are indicative.

\FloatBarrier

\section{Conclusion}\label{sec:conclusion}

We proved that exact edge cover by non-strict temporal walks is polynomially solvable for prescribed start counts, through M-convexity after removing an even component's dead state, and that the strict variants are NP-complete even with prescribed end vertices. The bound $O(D(n+sm))$ of Theorem~\ref{thm:time} is an upper bound on growth. On the shortage family the measured growth is slightly above $m^2$, and certificates from the augmentation loop rest on Lemma~\ref{lem:gamma}(b) and Lemma~\ref{lem:localmax}. We made no comparison with Boolean satisfiability or a generic submodular flow solver. Future work includes minimizing the number of walks with free starts, strict walks with at most two edges per vertex in each snapshot, and the dependence of $D$ on the initial flow.

\backmatter

\section*{Declarations}
\begin{itemize}
\item \textbf{Funding} None.
\item \textbf{Competing interests} None.
\item \textbf{Ethics approval and consent for publication} Not applicable. The Copenhagen Networks Study data are pseudonymised by the providers and used as released, under the providers' terms. No attempt was made to identify participants, and derived instance files are not redistributed.
\item \textbf{Use of AI assistants} The authors used AI assistants (Claude, Anthropic, and Codex, OpenAI) for language editing and translation, for drafting and checking parts of the text and proofs, for writing and testing code and experiment scripts, and for simulated peer review. The authors checked the results and take full responsibility for the content.
\item \textbf{Data availability} The Copenhagen Networks Study Bluetooth, call and SMS data~\cite{sapiezynski2019}, the SocioPatterns primary-school, high-school and hospital contact data~\cite{stehle2011,fournet2014,mastrandrea2015,vanhems2013} and the transit timetables~\cite{kujala2018} are available from their providers. Instances can be regenerated with the repository scripts.
\item \textbf{Code availability} \url{https://github.com/th1enq/exact-edge-cover-temporal-walks} contains the implementations, experiment scripts, generators and raw-result summaries. An archived version matching the reported results will be provided at acceptance.
\item \textbf{Author contributions} D.T.Q.: conceptualization, methodology, formal analysis, software, investigation, validation, original draft. M.C.T.: supervision, review and editing. Corresponding author.
\end{itemize}

\bibliography{references}

\begin{thebibliography}{32}
\ifx \bisbn   \undefined \def \bisbn  #1{ISBN #1}\fi
\ifx \binits  \undefined \def \binits#1{#1}\fi
\ifx \bauthor  \undefined \def \bauthor#1{#1}\fi
\ifx \batitle  \undefined \def \batitle#1{#1}\fi
\ifx \bjtitle  \undefined \def \bjtitle#1{#1}\fi
\ifx \bvolume  \undefined \def \bvolume#1{\textbf{#1}}\fi
\ifx \byear  \undefined \def \byear#1{#1}\fi
\ifx \bissue  \undefined \def \bissue#1{#1}\fi
\ifx \bfpage  \undefined \def \bfpage#1{#1}\fi
\ifx \blpage  \undefined \def \blpage #1{#1}\fi
\ifx \burl  \undefined \def \burl#1{\textsf{#1}}\fi
\ifx \doiurl  \undefined \def \doiurl#1{\url{https://doi.org/#1}}\fi
\ifx \betal  \undefined \def \betal{\textit{et al.}}\fi
\ifx \binstitute  \undefined \def \binstitute#1{#1}\fi
\ifx \binstitutionaled  \undefined \def \binstitutionaled#1{#1}\fi
\ifx \bctitle  \undefined \def \bctitle#1{#1}\fi
\ifx \beditor  \undefined \def \beditor#1{#1}\fi
\ifx \bpublisher  \undefined \def \bpublisher#1{#1}\fi
\ifx \bbtitle  \undefined \def \bbtitle#1{#1}\fi
\ifx \bedition  \undefined \def \bedition#1{#1}\fi
\ifx \bseriesno  \undefined \def \bseriesno#1{#1}\fi
\ifx \blocation  \undefined \def \blocation#1{#1}\fi
\ifx \bsertitle  \undefined \def \bsertitle#1{#1}\fi
\ifx \bsnm \undefined \def \bsnm#1{#1}\fi
\ifx \bsuffix \undefined \def \bsuffix#1{#1}\fi
\ifx \bparticle \undefined \def \bparticle#1{#1}\fi
\ifx \barticle \undefined \def \barticle#1{#1}\fi
\bibcommenthead
\ifx \bconfdate \undefined \def \bconfdate #1{#1}\fi
\ifx \botherref \undefined \def \botherref #1{#1}\fi
\ifx \url \undefined \def \url#1{\textsf{#1}}\fi
\ifx \bchapter \undefined \def \bchapter#1{#1}\fi
\ifx \bbook \undefined \def \bbook#1{#1}\fi
\ifx \bcomment \undefined \def \bcomment#1{#1}\fi
\ifx \oauthor \undefined \def \oauthor#1{#1}\fi
\ifx \citeauthoryear \undefined \def \citeauthoryear#1{#1}\fi
\ifx \endbibitem  \undefined \def \endbibitem {}\fi
\ifx \bconflocation  \undefined \def \bconflocation#1{#1}\fi
\ifx \arxivurl  \undefined \def \arxivurl#1{\textsf{#1}}\fi
\csname PreBibitemsHook\endcsname

\bibitem[\protect\citeauthoryear{Holme and Saram{\"a}ki}{2012}]{holme2012}
\begin{barticle}
\bauthor{\bsnm{Holme}, \binits{P.}},
\bauthor{\bsnm{Saram{\"a}ki}, \binits{J.}}:
\batitle{Temporal networks}.
\bjtitle{Physics Reports}
\bvolume{519}(\bissue{3}),
\bfpage{97}--\blpage{125}
(\byear{2012})
\doiurl{10.1016/j.physrep.2012.03.001}
\end{barticle}
\endbibitem

\bibitem[\protect\citeauthoryear{Michail}{2016}]{michail2016}
\begin{barticle}
\bauthor{\bsnm{Michail}, \binits{O.}}:
\batitle{An introduction to temporal graphs: an algorithmic perspective}.
\bjtitle{Internet Mathematics}
\bvolume{12}(\bissue{4}),
\bfpage{239}--\blpage{280}
(\byear{2016})
\doiurl{10.1080/15427951.2016.1177801}
\end{barticle}
\endbibitem

\bibitem[\protect\citeauthoryear{Edmonds and Johnson}{1973}]{edmonds1973}
\begin{barticle}
\bauthor{\bsnm{Edmonds}, \binits{J.}},
\bauthor{\bsnm{Johnson}, \binits{E.L.}}:
\batitle{Matching, {E}uler tours and the {C}hinese postman}.
\bjtitle{Mathematical Programming}
\bvolume{5}(\bissue{1}),
\bfpage{88}--\blpage{124}
(\byear{1973})
\doiurl{10.1007/BF01580113}
\end{barticle}
\endbibitem

\bibitem[\protect\citeauthoryear{Corber{\'a}n and Laporte}{2015}]{corberan2015}
\begin{bbook}
\beditor{\bsnm{Corber{\'a}n}, \binits{{\'A}.}},
\beditor{\bsnm{Laporte}, \binits{G.}} (eds.):
\bbtitle{Arc Routing: Problems, Methods, and Applications}.
\bsertitle{MOS-SIAM Series on Optimization},
vol. \bseriesno{20}.
\bpublisher{SIAM},
\blocation{Philadelphia}
(\byear{2015}).
\doiurl{10.1137/1.9781611973679}
\end{bbook}
\endbibitem

\bibitem[\protect\citeauthoryear{Bunte and Kliewer}{2009}]{bunte2009}
\begin{barticle}
\bauthor{\bsnm{Bunte}, \binits{S.}},
\bauthor{\bsnm{Kliewer}, \binits{N.}}:
\batitle{An overview on vehicle scheduling models}.
\bjtitle{Public Transport}
\bvolume{1}(\bissue{4}),
\bfpage{299}--\blpage{317}
(\byear{2009})
\doiurl{10.1007/s12469-010-0018-5}
\end{barticle}
\endbibitem

\bibitem[\protect\citeauthoryear{Casteigts et~al.}{2024}]{casteigts2024}
\begin{barticle}
\bauthor{\bsnm{Casteigts}, \binits{A.}},
\bauthor{\bsnm{Corsini}, \binits{T.}},
\bauthor{\bsnm{Sarkar}, \binits{W.}}:
\batitle{Simple, strict, proper, happy: a study of reachability in temporal
  graphs}.
\bjtitle{Theoretical Computer Science}
\bvolume{991},
\bfpage{114434}
(\byear{2024})
\doiurl{10.1016/j.tcs.2024.114434}
\end{barticle}
\endbibitem

\bibitem[\protect\citeauthoryear{Erlebach and Spooner}{2023}]{erlebach2023}
\begin{barticle}
\bauthor{\bsnm{Erlebach}, \binits{T.}},
\bauthor{\bsnm{Spooner}, \binits{J.T.}}:
\batitle{Parameterised temporal exploration problems}.
\bjtitle{Journal of Computer and System Sciences}
\bvolume{135},
\bfpage{73}--\blpage{88}
(\byear{2023})
\doiurl{10.1016/j.jcss.2023.01.003}
\end{barticle}
\endbibitem

\bibitem[\protect\citeauthoryear{Deligkas et~al.}{2025}]{deligkas2025}
\begin{bchapter}
\bauthor{\bsnm{Deligkas}, \binits{A.}},
\bauthor{\bsnm{D{\"o}ring}, \binits{M.}},
\bauthor{\bsnm{Eiben}, \binits{E.}},
\bauthor{\bsnm{Goldsmith}, \binits{T.-L.}},
\bauthor{\bsnm{Skretas}, \binits{G.}},
\bauthor{\bsnm{Tennigkeit}, \binits{G.}}:
\bctitle{How many lines to paint the city: Exact edge-cover in temporal
  graphs}.
In: \bbtitle{Proceedings of the AAAI Conference on Artificial Intelligence},
vol. \bseriesno{39},
pp. \bfpage{26498}--\blpage{26506}
(\byear{2025}).
\doiurl{10.1609/aaai.v39i25.34850}
\end{bchapter}
\endbibitem

\bibitem[\protect\citeauthoryear{Deligkas et~al.}{2024}]{deligkas2024arxiv}
\begin{botherref}
\oauthor{\bsnm{Deligkas}, \binits{A.}},
\oauthor{\bsnm{D{\"o}ring}, \binits{M.}},
\oauthor{\bsnm{Eiben}, \binits{E.}},
\oauthor{\bsnm{Goldsmith}, \binits{T.-L.}},
\oauthor{\bsnm{Skretas}, \binits{G.}},
\oauthor{\bsnm{Tennigkeit}, \binits{G.}}:
How Many Lines to Paint the City: Exact Edge-Cover in Temporal Graphs.
arXiv:2408.17107v1.
Version 1, submitted 30 August 2024
(2024).
\doiurl{10.48550/arXiv.2408.17107}
\end{botherref}
\endbibitem

\bibitem[\protect\citeauthoryear{Akrida et~al.}{2019}]{akrida2019}
\begin{barticle}
\bauthor{\bsnm{Akrida}, \binits{E.C.}},
\bauthor{\bsnm{Czyzowicz}, \binits{J.}},
\bauthor{\bsnm{G{\k{a}}sieniec}, \binits{L.}},
\bauthor{\bsnm{Kuszner}, \binits{{\L}.}},
\bauthor{\bsnm{Spirakis}, \binits{P.G.}}:
\batitle{Temporal flows in temporal networks}.
\bjtitle{Journal of Computer and System Sciences}
\bvolume{103},
\bfpage{46}--\blpage{60}
(\byear{2019})
\doiurl{10.1016/j.jcss.2019.02.003}
\end{barticle}
\endbibitem

\bibitem[\protect\citeauthoryear{Marino and Silva}{2023}]{marino2023}
\begin{barticle}
\bauthor{\bsnm{Marino}, \binits{A.}},
\bauthor{\bsnm{Silva}, \binits{A.}}:
\batitle{{E}ulerian walks in temporal graphs}.
\bjtitle{Algorithmica}
\bvolume{85}(\bissue{3}),
\bfpage{805}--\blpage{830}
(\byear{2023})
\doiurl{10.1007/s00453-022-01021-y}
\end{barticle}
\endbibitem

\bibitem[\protect\citeauthoryear{Bumpus and Meeks}{2023}]{bumpus2023}
\begin{barticle}
\bauthor{\bsnm{Bumpus}, \binits{B.M.}},
\bauthor{\bsnm{Meeks}, \binits{K.}}:
\batitle{Edge exploration of temporal graphs}.
\bjtitle{Algorithmica}
\bvolume{85}(\bissue{3}),
\bfpage{688}--\blpage{716}
(\byear{2023})
\doiurl{10.1007/s00453-022-01018-7}
\end{barticle}
\endbibitem

\bibitem[\protect\citeauthoryear{Chakraborty et~al.}{2024}]{chakraborty2024}
\begin{bchapter}
\bauthor{\bsnm{Chakraborty}, \binits{D.}},
\bauthor{\bsnm{Dailly}, \binits{A.}},
\bauthor{\bsnm{Foucaud}, \binits{F.}},
\bauthor{\bsnm{Klasing}, \binits{R.}}:
\bctitle{Algorithms and complexity for path covers of temporal {DAG}s}.
In: \bbtitle{49th International Symposium on Mathematical Foundations of
  Computer Science (MFCS 2024)}.
\bsertitle{LIPIcs},
vol. \bseriesno{306},
pp. \bfpage{38}--\blpage{13817}
(\byear{2024}).
\doiurl{10.4230/LIPIcs.MFCS.2024.38}
\end{bchapter}
\endbibitem

\bibitem[\protect\citeauthoryear{Murota}{1996}]{murota1996}
\begin{barticle}
\bauthor{\bsnm{Murota}, \binits{K.}}:
\batitle{Convexity and steinitz's exchange property}.
\bjtitle{Advances in Mathematics}
\bvolume{124}(\bissue{2}),
\bfpage{272}--\blpage{310}
(\byear{1996})
\doiurl{10.1006/aima.1996.0084}
\end{barticle}
\endbibitem

\bibitem[\protect\citeauthoryear{Murota}{2003}]{murota2003}
\begin{bbook}
\bauthor{\bsnm{Murota}, \binits{K.}}:
\bbtitle{Discrete Convex Analysis}.
\bsertitle{SIAM Monographs on Discrete Mathematics and Applications},
vol. \bseriesno{10}.
\bpublisher{SIAM},
\blocation{Philadelphia}
(\byear{2003}).
\doiurl{10.1137/1.9780898718508}
\end{bbook}
\endbibitem

\bibitem[\protect\citeauthoryear{Edmonds and Giles}{1977}]{edmonds1977}
\begin{bchapter}
\bauthor{\bsnm{Edmonds}, \binits{J.}},
\bauthor{\bsnm{Giles}, \binits{R.}}:
\bctitle{A min-max relation for submodular functions on graphs}.
In: \beditor{\bsnm{Hammer}, \binits{P.L.}},
\beditor{\bsnm{Johnson}, \binits{E.L.}},
\beditor{\bsnm{Korte}, \binits{B.H.}} (eds.)
\bbtitle{Studies in Integer Programming}.
\bsertitle{Annals of Discrete Mathematics},
vol. \bseriesno{1},
pp. \bfpage{185}--\blpage{204}.
\bpublisher{North-Holland},
\blocation{Amsterdam}
(\byear{1977}).
\doiurl{10.1016/S0167-5060(08)70734-9}
\end{bchapter}
\endbibitem

\bibitem[\protect\citeauthoryear{Frank and Tardos}{1988}]{franktardos1988}
\begin{barticle}
\bauthor{\bsnm{Frank}, \binits{A.}},
\bauthor{\bsnm{Tardos}, \binits{{\'E}.}}:
\batitle{Generalized polymatroids and submodular flows}.
\bjtitle{Mathematical Programming}
\bvolume{42},
\bfpage{489}--\blpage{563}
(\byear{1988})
\doiurl{10.1007/BF01589418}
\end{barticle}
\endbibitem

\bibitem[\protect\citeauthoryear{Fujishige}{2005}]{fujishige2005}
\begin{bbook}
\bauthor{\bsnm{Fujishige}, \binits{S.}}:
\bbtitle{Submodular Functions and Optimization},
\bedition{2nd} edn.
\bsertitle{Annals of Discrete Mathematics},
vol. \bseriesno{58}.
\bpublisher{Elsevier},
\blocation{Amsterdam}
(\byear{2005})
\end{bbook}
\endbibitem

\bibitem[\protect\citeauthoryear{Hakimi}{1965}]{hakimi1965}
\begin{barticle}
\bauthor{\bsnm{Hakimi}, \binits{S.L.}}:
\batitle{On the degrees of the vertices of a directed graph}.
\bjtitle{Journal of the Franklin Institute}
\bvolume{279}(\bissue{4}),
\bfpage{290}--\blpage{308}
(\byear{1965})
\doiurl{10.1016/0016-0032(65)90340-6}
\end{barticle}
\endbibitem

\bibitem[\protect\citeauthoryear{Frank}{1980}]{frank1980}
\begin{barticle}
\bauthor{\bsnm{Frank}, \binits{A.}}:
\batitle{On the orientation of graphs}.
\bjtitle{Journal of Combinatorial Theory, Series B}
\bvolume{28}(\bissue{3}),
\bfpage{251}--\blpage{261}
(\byear{1980})
\doiurl{10.1016/0095-8956(80)90071-4}
\end{barticle}
\endbibitem

\bibitem[\protect\citeauthoryear{Frank}{2011}]{frank2011}
\begin{bbook}
\bauthor{\bsnm{Frank}, \binits{A.}}:
\bbtitle{Connections in Combinatorial Optimization}.
\bsertitle{Oxford Lecture Series in Mathematics and its Applications},
vol. \bseriesno{38}.
\bpublisher{Oxford University Press},
\blocation{Oxford}
(\byear{2011})
\end{bbook}
\endbibitem

\bibitem[\protect\citeauthoryear{Cunningham and Frank}{1985}]{cunningham1985}
\begin{barticle}
\bauthor{\bsnm{Cunningham}, \binits{W.H.}},
\bauthor{\bsnm{Frank}, \binits{A.}}:
\batitle{A primal-dual algorithm for submodular flows}.
\bjtitle{Mathematics of Operations Research}
\bvolume{10}(\bissue{2}),
\bfpage{251}--\blpage{262}
(\byear{1985})
\doiurl{10.1287/moor.10.2.251}
\end{barticle}
\endbibitem

\bibitem[\protect\citeauthoryear{Abdi et~al.}{2026}]{abdi2025}
\begin{barticle}
\bauthor{\bsnm{Abdi}, \binits{A.}},
\bauthor{\bsnm{Cornu{\'e}jols}, \binits{G.}},
\bauthor{\bsnm{Liu}, \binits{S.}},
\bauthor{\bsnm{Silina}, \binits{O.}}:
\batitle{Strongly connected orientations and integer lattices}.
\bjtitle{Mathematics of Operations Research}
(\byear{2026})
\doiurl{10.1287/moor.2025.1030}
{\href{https://arxiv.org/abs/2410.13665}{{arXiv:2410.13665}}}.
\bcomment{Published online 3 August 2026; a conference version appeared at IPCO
  2025}
\end{barticle}
\endbibitem

\bibitem[\protect\citeauthoryear{Hierholzer}{1873}]{hierholzer1873}
\begin{barticle}
\bauthor{\bsnm{Hierholzer}, \binits{C.}}:
\batitle{Ueber die {M}{\"o}glichkeit, einen {L}inienzug ohne {W}iederholung und
  ohne {U}nterbrechung zu umfahren}.
\bjtitle{Mathematische Annalen}
\bvolume{6},
\bfpage{30}--\blpage{32}
(\byear{1873})
\doiurl{10.1007/BF01442866}
\end{barticle}
\endbibitem

\bibitem[\protect\citeauthoryear{Tovey}{1984}]{tovey1984}
\begin{barticle}
\bauthor{\bsnm{Tovey}, \binits{C.A.}}:
\batitle{A simplified {NP}-complete satisfiability problem}.
\bjtitle{Discrete Applied Mathematics}
\bvolume{8}(\bissue{1}),
\bfpage{85}--\blpage{89}
(\byear{1984})
\doiurl{10.1016/0166-218X(84)90081-7}
\end{barticle}
\endbibitem

\bibitem[\protect\citeauthoryear{Huangfu and Hall}{2018}]{huangfu2018}
\begin{barticle}
\bauthor{\bsnm{Huangfu}, \binits{Q.}},
\bauthor{\bsnm{Hall}, \binits{J.A.J.}}:
\batitle{Parallelizing the dual revised simplex method}.
\bjtitle{Mathematical Programming Computation}
\bvolume{10}(\bissue{1}),
\bfpage{119}--\blpage{142}
(\byear{2018})
\doiurl{10.1007/s12532-017-0130-5}
\end{barticle}
\endbibitem

\bibitem[\protect\citeauthoryear{Sapiezynski et~al.}{2019}]{sapiezynski2019}
\begin{barticle}
\bauthor{\bsnm{Sapiezynski}, \binits{P.}},
\bauthor{\bsnm{Stopczynski}, \binits{A.}},
\bauthor{\bsnm{Lassen}, \binits{D.D.}},
\bauthor{\bsnm{Lehmann}, \binits{S.}}:
\batitle{Interaction data from the {C}openhagen {N}etworks {S}tudy}.
\bjtitle{Scientific Data}
\bvolume{6},
\bfpage{315}
(\byear{2019})
\doiurl{10.1038/s41597-019-0325-x}
\end{barticle}
\endbibitem

\bibitem[\protect\citeauthoryear{Stehl{\'e} et~al.}{2011}]{stehle2011}
\begin{barticle}
\bauthor{\bsnm{Stehl{\'e}}, \binits{J.}},
\bauthor{\bsnm{Voirin}, \binits{N.}},
\bauthor{\bsnm{Barrat}, \binits{A.}},
\bauthor{\bsnm{Cattuto}, \binits{C.}},
\bauthor{\bsnm{Isella}, \binits{L.}},
\bauthor{\bsnm{Pinton}, \binits{J.-F.}},
\bauthor{\bsnm{Quaggiotto}, \binits{M.}},
\bauthor{\bsnm{Broeck}, \binits{W.}},
\bauthor{\bsnm{R{\'e}gis}, \binits{C.}},
\bauthor{\bsnm{Lina}, \binits{B.}},
\bauthor{\bsnm{Vanhems}, \binits{P.}}:
\batitle{High-resolution measurements of face-to-face contact patterns in a
  primary school}.
\bjtitle{PLoS ONE}
\bvolume{6}(\bissue{8}),
\bfpage{23176}
(\byear{2011})
\doiurl{10.1371/journal.pone.0023176}
\end{barticle}
\endbibitem

\bibitem[\protect\citeauthoryear{Fournet and Barrat}{2014}]{fournet2014}
\begin{barticle}
\bauthor{\bsnm{Fournet}, \binits{J.}},
\bauthor{\bsnm{Barrat}, \binits{A.}}:
\batitle{Contact patterns among high school students}.
\bjtitle{PLoS ONE}
\bvolume{9}(\bissue{9}),
\bfpage{107878}
(\byear{2014})
\doiurl{10.1371/journal.pone.0107878}
\end{barticle}
\endbibitem

\bibitem[\protect\citeauthoryear{Mastrandrea et~al.}{2015}]{mastrandrea2015}
\begin{barticle}
\bauthor{\bsnm{Mastrandrea}, \binits{R.}},
\bauthor{\bsnm{Fournet}, \binits{J.}},
\bauthor{\bsnm{Barrat}, \binits{A.}}:
\batitle{Contact patterns in a high school: a comparison between data collected
  using wearable sensors, contact diaries and friendship surveys}.
\bjtitle{PLoS ONE}
\bvolume{10}(\bissue{9}),
\bfpage{0136497}
(\byear{2015})
\doiurl{10.1371/journal.pone.0136497}
\end{barticle}
\endbibitem

\bibitem[\protect\citeauthoryear{Vanhems et~al.}{2013}]{vanhems2013}
\begin{barticle}
\bauthor{\bsnm{Vanhems}, \binits{P.}},
\bauthor{\bsnm{Barrat}, \binits{A.}},
\bauthor{\bsnm{Cattuto}, \binits{C.}},
\bauthor{\bsnm{Pinton}, \binits{J.-F.}},
\bauthor{\bsnm{Khanafer}, \binits{N.}},
\bauthor{\bsnm{R{\'e}gis}, \binits{C.}},
\bauthor{\bsnm{Kim}, \binits{B.-a.}},
\bauthor{\bsnm{Comte}, \binits{B.}},
\bauthor{\bsnm{Voirin}, \binits{N.}}:
\batitle{Estimating potential infection transmission routes in hospital wards
  using wearable proximity sensors}.
\bjtitle{PLoS ONE}
\bvolume{8}(\bissue{9}),
\bfpage{73970}
(\byear{2013})
\doiurl{10.1371/journal.pone.0073970}
\end{barticle}
\endbibitem

\bibitem[\protect\citeauthoryear{Kujala et~al.}{2018}]{kujala2018}
\begin{barticle}
\bauthor{\bsnm{Kujala}, \binits{R.}},
\bauthor{\bsnm{Weckstr{\"o}m}, \binits{C.}},
\bauthor{\bsnm{Darst}, \binits{R.K.}},
\bauthor{\bsnm{Mladenovi{\'c}}, \binits{M.N.}},
\bauthor{\bsnm{Saram{\"a}ki}, \binits{J.}}:
\batitle{A collection of public transport network data sets for 25 cities}.
\bjtitle{Scientific Data}
\bvolume{5},
\bfpage{180089}
(\byear{2018})
\doiurl{10.1038/sdata.2018.89}
\end{barticle}
\endbibitem

\end{thebibliography}

\end{document}